\documentclass[aps,twocolumn,amsmath,amssymb,a4paper,9pt,showkeys,notitlepage,nofootinbib]{revtex4}
\def\pdftitle{U-turn}
\def\authorname{Marc Pichel}
\def\pdfsubject{}
\def\pdfkeywords{}
\def\pdfbackref{none}

\usepackage[pdftex,
        colorlinks=true,
        urlcolor=darkblue,               % \href{...}{...}
        anchorcolor=darkblue,
        filecolor=green,                 % \href*{...}
        linkcolor=darkblue,              % \ref{...} and \pageref{...}
        menucolor=darkblue,
        citecolor=darkblue,
        pagebackref,
        backref={\pdfbackref},
        pdfpagemode={UseOutlines},
        bookmarks=true,
        bookmarksopen=true,
        pdftitle={\pdftitle},
        pdfauthor={\authorname}, 
        pdfsubject={\pdfsubject},
        pdfkeywords={\pdfkeywords},
        ]{hyperref}

\usepackage{graphicx}
\DeclareGraphicsExtensions{.pdf,.png,.jpg}
\usepackage{thumbpdf}
\usepackage{color}
\definecolor{darkgreen}{rgb}{0,0.5,0}
\definecolor{darkblue}{rgb}{0,0,0.5}
\definecolor{brown}{rgb}{0.98,0.92,0.73}
\definecolor{red}{rgb}{1,0,0}
\definecolor{yellow}{rgb}{1,1,0}
\definecolor{blue}{rgb}{0,0,1}
\definecolor{green}{rgb}{0,1,0}
\definecolor{purple}{rgb}{1,0,1}
\definecolor{gray}{rgb}{0.8,0.8,0.8}
\definecolor{black}{rgb}{0,0,0}
\definecolor{white}{rgb}{1,1,1}
\definecolor{gold}{rgb}{1.,0.84,0.}

\usepackage{amsmath}
\usepackage{wasysym}
\usepackage{booktabs}

\usepackage{lineno}

\usepackage{pdfsync}

\usepackage{doi}

\usepackage{siunitx}

\usepackage{subfig}

\usepackage{xspace} %The \xspace command adds space at the end of a macro designed for use in text, unless the macro is followed by certain punctuation characters.
\def\um{$\mu$m\xspace}
\def\cm2{cm$^2$\xspace}

\def\degree{$^o$\xspace}

\usepackage{amsmath}
\def\bs{\boldsymbol}
\newif\ifmac
\mactrue

\def\leonmovie[#1]#2#3#4#5{%
  \ifmac%Image is clickable
  \begin{figure}
    \centering
    \easymovie[#1]{
      \includegraphics[keepaspectratio, width=#2, height=#3]
      {#4}}
    {#5}
  \end{figure}
  \else%Link below image
  \begin{figure}
    \centering
    \includegraphics[keepaspectratio, width=#2, height=#3-\lineheight]
    {#4}
  \end{figure}
  \begin{center}
    \begin{small}
    \href{run:#5}{Movie}
    \end{small}
  \end{center}
  \fi
}

\usepackage[paper=a4paper,total={170mm,241mm},top=20mm,left=20mm,columnsep=8mm]{geometry}
\usepackage{siunitx}
\usepackage{nicefrac}

\def\narrowfigurewidth{0.6\columnwidth}
\def\figurewidth{0.8\columnwidth}
\def\widefigurewidth{\columnwidth}

\usepackage{color}
\usepackage{subfig} 

\newif\ifcmtr
\cmtrtrue
\ifcmtr %  Highlight comment (and hide it when necessary)
\newcommand{\cmtr}[1]{ %
   [\color{red} \textbf{#1} \normalcolor]%
}%
\else
\newcommand{\cmtr}[1]{ %
}% 
\fi

\newif\ifcmtrla
\cmtrlatrue
\ifcmtr %  Highlight comment (and hide it when necessary)
\newcommand{\cmtrla}[1]{ %
   [\color{blue} \textbf{#1} \normalcolor]%
}%
\else
\newcommand{\cmtrla}[1]{ %
}%
\fi

\newif\ifcmtrth
\cmtrthtrue
\ifcmtr %  Highlight comment (and hide it when necessary)
\newcommand{\cmtrth}[1]{ %
   [\color{magenta} \textbf{#1} \normalcolor]%
}%
\else
\newcommand{\cmtrth}[1]{ %
}%
\fi

\newif\ifcmtrmp
\cmtrmptrue
\ifcmtr %  Highlight comment (and hide it when necessary)
\newcommand{\cmtrmp}[1]{ %
   [\color{red} \textbf{#1} \normalcolor]%
}%
\else
\newcommand{\cmtrmp}[1]{ %
}%
\fi

\begin{document}

\title{Magnetic Response of \emph{Magnetospirillum Gryphiswaldense}}
% research question: How do magneto tactic bacteria make a U-turn
\author{M. P. Pichel$^{1,2}$}
\author{T. A. G. Hageman$^{1,2}$}
\author{I. S. M. Khalil$^{3}$}
% \author{S. Misra$^{4}$} 
\author{A. Manz$^{1}$} 
\author{L. Abelmann$^{1,2}$} 
\affiliation{$^{1}$KIST Europe, Saarbr\"ucken, Germany, \\
$^{2}$MIRA$^+$ Research Institute, University of Twente, The Netherlands\\
$^{3}$The German University in Cairo, New Cairo City, Egypt 
\underline{l.abelmann@kist-europe.de}}
%\date{\today}
\vskip10mm
\begin{abstract}
  In this study we modelled and measured the U-turn trajectories of
  individual magnetotactic bacteria under the application of rotating
  magnetic fields, ranging in ampitude from 1 to \SI{12}{mT}. The
  model is based on the balance between rotational drag and magnetic
  torque. For accurate verification of this model, bacteria were
  observed inside \SI{5}{\um} tall microfluidic channels, so that they
  remained in focus during the entire trajectory. From the analysis of
  hundreds of trajectories and accurate measurements of bacteria and
  magnetosome chain dimensions, we confirmed that the model is correct
  within measurement error. The resulting average rate of
  rotation of \emph{Magnetospirillum Gryphiswaldense} is \SI{0.74(3)}{rad/mTs}.
\end{abstract}


\maketitle
% \makeatletter 
% \renewcommand{\@tocrmarg}{2.55em plus1fil} 
% \renewcommand{\@pnumwidth}{3em} 
% \renewcommand{\@tocrmarg}{4em} 
% \makeatother

%\tableofcontents

\section{Introduction}
\label{sec:introduction}
% wat was, wat is, wie deed wat, wat doen wij, etc.

%\subsection{Research Question}
Magnetotactic bacteria (MTB\footnote{Throughout this paper we will use
  the acronym
  MTB to indicate the single bacterium as well as multiple bacteria}) possess an internal chain of magnetosome vesicles~\cite{Komeili2004} which
biomineralise nanometer sized magnetic crystals (Fe$_{3}$O$_{4}$ or Fe$_{3}$S$_{4}$~\cite{
  Lins2005, Baumgartner2011, Uebe2016}), encompassed by a
 membrane (magnetosome)~\cite{Gorby1988}. This magnetosome chain (MC) acts
much like a compass needle. The magnetic torque acting on the
MC aligns the bacteria with the earth magnetic
field~\cite{Erglis2007}. This is a form of
magnetoception~\cite{Kirschvink2001}, working in conjunction with
aero-taxis~\cite{Frankel1997}.  At high latitudes the earth's magnetic
field is not only aligned North-South, but also substantially inclined
with respect to the earth's surface~\cite{Maus2010}. The MTB are
therefore aligned vertically, which converts a three-dimensional
search for the optimal (oxygen) conditions into a more efficient
one-dimensional search~\cite{Esquivel1986} (gravitational forces do
not play a significant role at the scale of a bacterium). This gives MTB an
evolutionary advantage over non-magnetic bacteria in environments with
stationary chemical gradients more or less perpendicular to the water
surface.

In this paper we address the question of how the MTB of type
\emph{Magnetosprilillum Gryphiswaldense} (MSR-1) respond to varying
magnitudes of the external field, in particular a field that is rotating.
%\subsection{Why is this question relevant?}
Even though the response of individual magneto-tactic bacteria to an
external magnetic field has been
modelled and observed~\cite{Bahaj1993,Bahaj1996,Kampen1995, Erglis2007,Cebers2011},
there has been no thorough observation of the dependence on the field
strength. The existing models predict a linear relation between
the angular velocity of the bacterium and the field strength, but this has
not been confirmed experimentally. Nor has there been an analysis of the spread
in response over the population of bacteria. The main reason for the
absence of experimental data is that the depth of focus at the
magnification required prohibits the observation of multiple bacteria in
parallel. In this paper, we introduce microfluidic chips with a channel depth
of only \SI{5}{\um}, which ensures that all bacteria in the field of
view remain in focus.

The second motivation for studying the response of MTB to external
magnetic fields, is that they are an ideal model system for self
propelled medical microrobotics~\cite{Menciassi2007, Abbott2009}.
Medical microrobotics is a novel form of minimally invasive surgery
(MIS), in which one tries to reduce the patient's surgical trauma while
enabling clinicians to reach deep seated locations within the human
body~\cite{Nelson2004, Abayazid2013, Felfoul2016}.

The current approach in medical microrobotics is to insert the
miniaturized tools needed for a medical procedure into the patient
through a small insertion or orifice. By reducing the size of these
tools a larger range of natural pathways becomes available. Currently,
these tools are mechanically connected to the outside world. If this
connection can be removed, so that the tools become untethered,
(autonomous) manoeuvring through the veins and arteries of the body
becomes possible~\cite{Dankelman2011}.

If the size and/or application of these untethered systems inside the
human body prohibits the storage of energy for propulsion, the energy has
to be harvested from the environment. One solution is the use of
alternating magnetic fields~\cite{Abbott2009}. This method is simple,
but although impressive progress has been made, it is appallingly
inefficient. Only a fraction, ~\num{e-12}, of the supplied energy
field is actually used by the microrobot. This is not a problem for
microscopy experiments, but will become a serious issue if the
microrobots are to be controlled deep inside the human
body. The efficiency would increase dramatically if the microrobot could
harvest its energy from the surrounding liquid. In human blood, energy
is abundant and used by all cells for respiration.

For self-propelled objects, only the direction of motion needs to be
controlled by the external magnetic field. There is no need for field
gradients to apply forces, so the field is allowed to be weaker and
uniform when solely using magnetic torque~\cite{Nelson2010}. Compared
to systems that derive their energy for propulsion from the magnetic
field, the field can be small in magnitude and only needs to vary slowly. As a result, the energy requirements are low and overheating problems can be avoided.

Nature provides us with a plenitude of self-propelled micro-organisms that derive their energy from bio-compatible liquids, as described first by Bellini~\cite{Bellini1963}. MTB provide a perfect biokleptic model to test concepts and study the behaviour of self-propelled micro-objects steered by external magnetic fields~\cite{Khalil2013b}.

%\subsection{Prior state-of-the-art}
The direction of the motion of an MTB is modified by the application of a
magnetic field at an angle with the easy axis of magnetization of
the magnetosome. The resulting magnetic torque causes a rotation of
the MTB at a speed that is determined by the balance between the
magnetic torque and the rotational drag torque. Under the application of a
uniform rotating field, the bacteria follow U-turn
trajectories~\cite{Bahaj1993, Yang2012, Reufer2014}.

The magnetic torque is often modelled by assuming that the magnetic
element is a permanent magnet with dipole moment $\bs{m}$ [Am$^2$] on
which the magnetic field $\bs{B}$ [T] exerts a torque
$\bs{\Gamma}=\bs{m}\times\bs{B}$ [Nm]. This simple model suggest that
the torque increases linearly with the field strength, where it is assumed that the
atomic dipoles are rigidly fixed to the lattice, and hardly rotate at
all. This is usually only the case for very small magnetic fields. 

In general one should consider a change in the magnetic energy as a function
of the magnetization direction with respect to the object (magnetic
anisotropy). This is correctly suggested by Erglis \emph{et al.} for
magnetotactic bacteria~\cite{Erglis2007}. An estimation of the
magnetic dipole moment can be obtained by studying the dynamics of
MTB~\cite{Bahaj1996}. 

Recent studies of the dynamics of MTB in a rotating
magnetic field show that a random walk is still present regardless of
the presence of a rotating field~\cite{Smid2015,
  Cebers2011}. The formation and control of aggregates of MTB in both two-
and three-dimensional control systems has been achieved \emph{in
  vitro}~\cite{Martel2009, Martel2010, DeLanauze2014} as well as
\emph{in vivo}~\cite{Felfoul2016}, showing that MTB can use the natural
hypoxic state surrounding cancerous tissue for targeted drug delivery.

Despite these impressive results, successful control of individual
MTB is much less reported. This is because many experiments suffer
from a limited depth of focus of the microscope system, leading to
a loss of tracking. A collateral problem is overheating of
the electromagnets in experiments that take longer than a few minutes. We
recently demonstrated the effect of varying field strengths on the
control of magneto-tactic bacteria~\cite{Hassan2016}. In the present paper we
provide the theoretical framework and systematically analyse the
influence of the magnetic field on the trajectories of individual MTB.
This knowledge will contribute to more efficient control of individual
MTB, and ultimately self-propelled robotic systems in general.

% Rommel
%because of their streerability, MTBs have been suggested for medical applications ranging from (bio)engineered micro~\cite{Nelson2010} and nano robots~\cite{Lenaghan2013} to cancer therapies~\cite{Forbes2010,  Dasdag2014, Mathuriya2014}, to environmental clean-up \cmtr{niet  echt medical} \cite{Bahaj1993b}, to general biocarriers~\cite{Zu2006} to controllable MRI trackers~\cite{Martel2009a}.

%\subsection{Organisation of paper}
We present a thorough theoretical analysis
of the magnetic and drag torques on MTB. This model is used to
derive values for the proportionality between the average rate of rotation and
the magnetic field during a U-turn trajectory under a magnetic field
reversal. The theory is used to predict U-turn trajectories of
MTB, which are the basis for our experimental procedures.

Lastely, we present statistically significant experimental results which verify our theoretical approach and employ a realistic range of magnetic field strength and rotational speed of the applied magnetic field to minimize energy input. 

%%% Local Variables:
%%% mode: latex
%%% TeX-master: "UTurn"
%%% End:
\section{Theory}
\label{sec:theory}

% subsectie Leon & Lars
\subsection{The Rate of Rotation}
\label{sec:angularvel}

\subsubsection{The dependence on the field}
The magnetic torque $\Gamma$ [Nm] is equal to the change in total
magnetic energy $U$ [J] with changing applied field angle. We consider
only the demagnetization and external field energy terms. The
demagnetization energy is caused by the magnetic stray field
$\bs{H}_\text{d}$ [A/m] that arises due to the magnetosome
magnetization $\bs{M}$ [A/m]. In principle, one has to integrate the
stray field over all space. Fortunately, this integral is
mathematically equivalent to~\cite{Hubert1998}

\begin{equation}
U_\text{d}=\textstyle{\frac{1}{2}}\mu_0\int\bs{M}\cdot\bs{H_\text{d}}\text{d}V, 
\end{equation}

\noindent with $\mu_o$ the vacuum permeability, $4\pi10^{-7}$. In this
formulation, the integral is conveniently restricted to the volume $V$
of the magnetic material.

The demagnetization energy acts to orient the magnetization so that
the external stray field energy is minimized. We can define a shape
anisotropy term $K$ [J/m$^3$] to represent the energy difference
between the hard and easy axes of magnetization, which are 
perpendicular to each other,

\begin{equation}
K=\left(U_\text{d, max}-U_\text{d, min}\right)/V.
\end{equation}

The external field energy is caused by the externally applied field $\bs{H}$ [A/m]

\begin{equation}
U_\text{H}=-\mu_0\int\bs{M}\cdot\bs{H}\text{d}V,
\end{equation}

\noindent and acts to align $\bs{M}$ parallel to $\bs{H}$. Assuming that the
magnetic element of volume $V$ is uniformly magnetized with saturation
magnetization $M_\text{s}$ [A/m], the total energy can then be 
expressed as

\begin{equation}
  U=KVsin^2(\theta)-\mu_0M_\text{s}HVcos(\varphi-\theta).
\end{equation}

\begin{figure}
  \begin{centering}
        \includegraphics[width=\narrowfigurewidth]{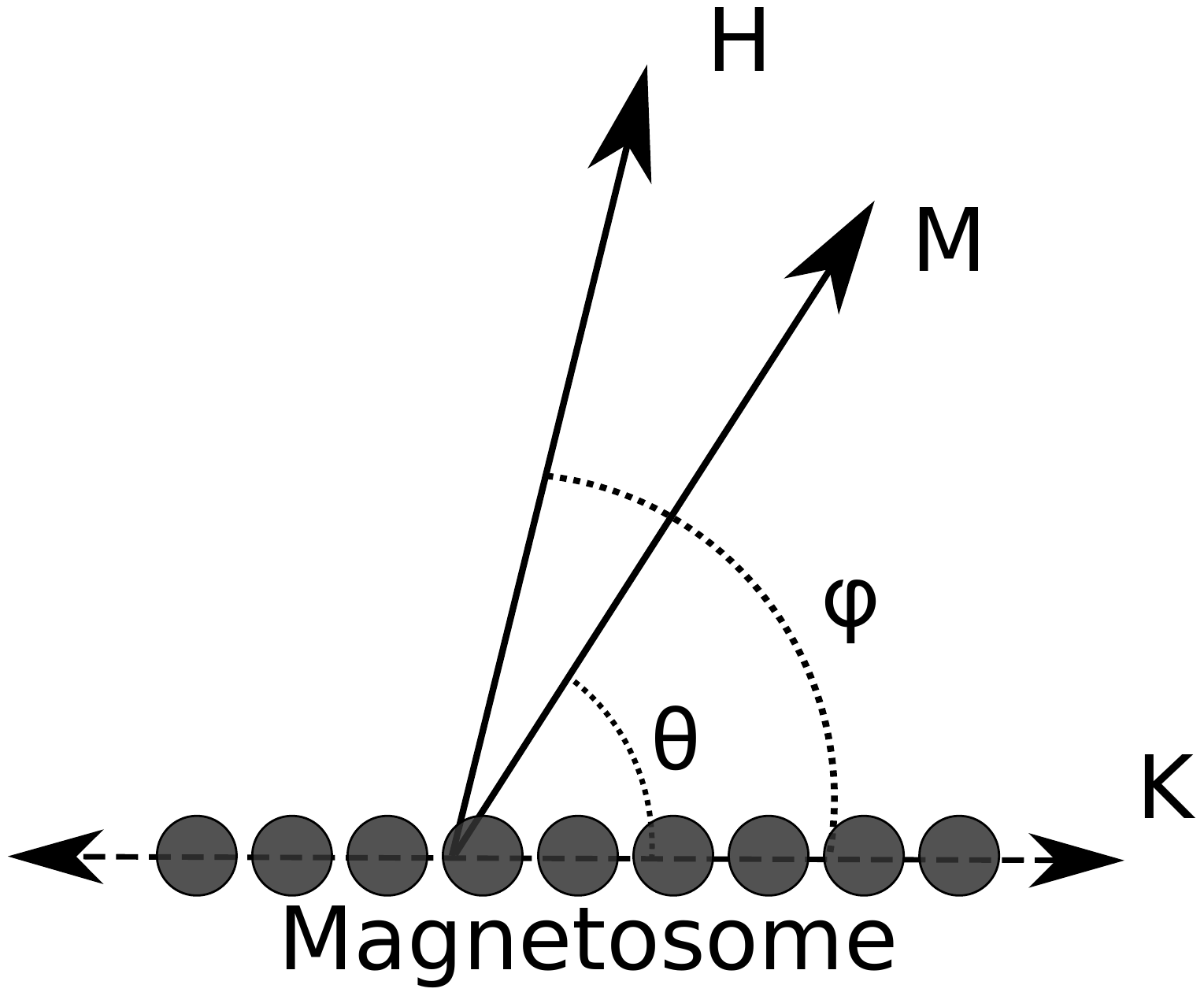}
  \end{centering}
  \caption{Definition of the field angle $\varphi$ and the magnetization angle
    $\theta$ between the easy axis $K$, the magnetization
    $M$ and the magnetic field $H$.}
  \label{fig:AxisMandH}
\end{figure}

The angles $\theta$ and $\varphi$ are defined as in
figure~\ref{fig:AxisMandH}. Normalizing the energy, field, and torque by

\begin{align}
u&=\nicefrac{U}{KV}\\
h&=\nicefrac{\mu_0HM}{2K}\label{eq:6}\\
\tau&=\nicefrac{\Gamma}{KV},
\end{align}

\noindent respectively, the expression for the energy can be simplified to

\begin{equation}
u=\sin^2(\theta)-2h\cos(\varphi-\theta).
\end{equation}

The equilibrium magnetization direction is reached for
$\nicefrac{\partial u}{\partial \theta}=0$. The solution for this
relationship cannot be expressed in an analytically concise
form. The main results are however that for
$h<\nicefrac{1}{\sqrt{2}}$, the maximum torque is reached at the field angle
$\varphi_\text{max}=\nicefrac{\pi}{2}$,

\begin{align}
 \tau_\text{max}&=2h\sqrt{1-h^2} &\text{ for } h\le\nicefrac{1}{\sqrt{2}}\label{eq:9}\\
  &=1 &\text{ for } h>\nicefrac{1}{\sqrt{2}}.
\end{align}

The angle of magnetization at maximum torque can be approximated by

\begin{equation}
\theta_\text{max} =h+0.1 h^2  \text{ for
} h< \nicefrac{1}{\sqrt{2}},\\
\end{equation}

\noindent where the error is smaller than $5\times10^{-3}$ rad (\SI{1.6}{\degree}) for $h<0.5$.

For $h>1$, the field angle $\varphi_\text{max}$ at which the maximum
torque is reached is smaller than $\nicefrac{\pi}{2}$ and approaches
$\nicefrac{\pi}{4}$ for $h\rightarrow\infty$. This behaviour can be very
well approximated by

\begin{align}
\varphi_\text{max}=\frac{\pi}{4}\left(1+\frac{2}{3h}\right) \text{ for
} h>1,
\end{align}

\noindent where the error is smaller than $3\times10^{-3}\pi$
(\SI{0.5}{\degree}).

In summary, and returning to variables with units, the maximum
torque is $\Gamma_\text{max}=KV $, which is reached at 

\begin{equation}
H>\frac{\sqrt{2}K}{\mu_0M_\text{s}}
\end{equation}

\noindent at an angle $\varphi=\nicefrac{\pi}{2}$, which, to a good approximation, decreases 
linearly with $1/H$ to
$\varphi=\nicefrac{\pi}{4}$ at an infinite external field. 

\subsubsection{Demagnetization factor}

The magnetization $M_\text{s}$ is a material parameter, so the only
variable to be determined is the magnetosome's demagnetization factor.
As a first approximation, we can consider the chain of magnetic crystals
in the magnetosome as a chain of $n$ dipoles separated at a distance
$a$, each with a dipole moment $m$=$M_sV$ [Am$^2$], where $V$ is the
volume of each single sphere. We assume that all dipoles are aligned
parallel to the field ($\varphi=\theta$) to obtain an upper limit on
the torque. (See figure~\ref{fig:AxisMandH} for the definition of the angles). The
magnetic energy for such a dipole chain has been derived by Jacobs and
Bean~\cite{Jacobs1955}, which, rewritten in SI units, is

\begin{align}
  U=&\frac{\mu_0m^2}{4\pi a^3} n K_n\left(1-3\cos^2(\theta)\right) + \nonumber\\
  &\mu_0nmHcos(\varphi-\theta)\\
 K_n=&\sum_{j=1}^n\frac{(n-j)}{nj^3}.
\end{align}

The maximum torque equals the energy difference between the state
where all moments are parallel to the chain ($\theta$=0) and the state
where they are perpendicular to the chain ($\theta$=$\pi/2$), under
the condition that the angle between the moments and the field is zero:

\begin{align}
  \label{eq:22}
  \Gamma_\text{max}&=\frac{3\mu_0m^2}{4\pi a^3} n K_n.
\end{align}

For a single dipole $n=1$, $K_n$=0 and there is no energy
difference, as expected. 

Combined with equations~(\ref{eq:6}) and (\ref{eq:9}), and re-introducing
units, the field dependent torque becomes

\begin{align}
  \label{eq:fielddependence}
  \Gamma&=\Gamma_\text{max}2h\sqrt{1-h^2}\text{ with }\\
  h&=\frac{H}{\Delta N M_\text{s}}.
\end{align}

\begin{figure}
  \begin{centering}
    \includegraphics[width=\narrowfigurewidth]{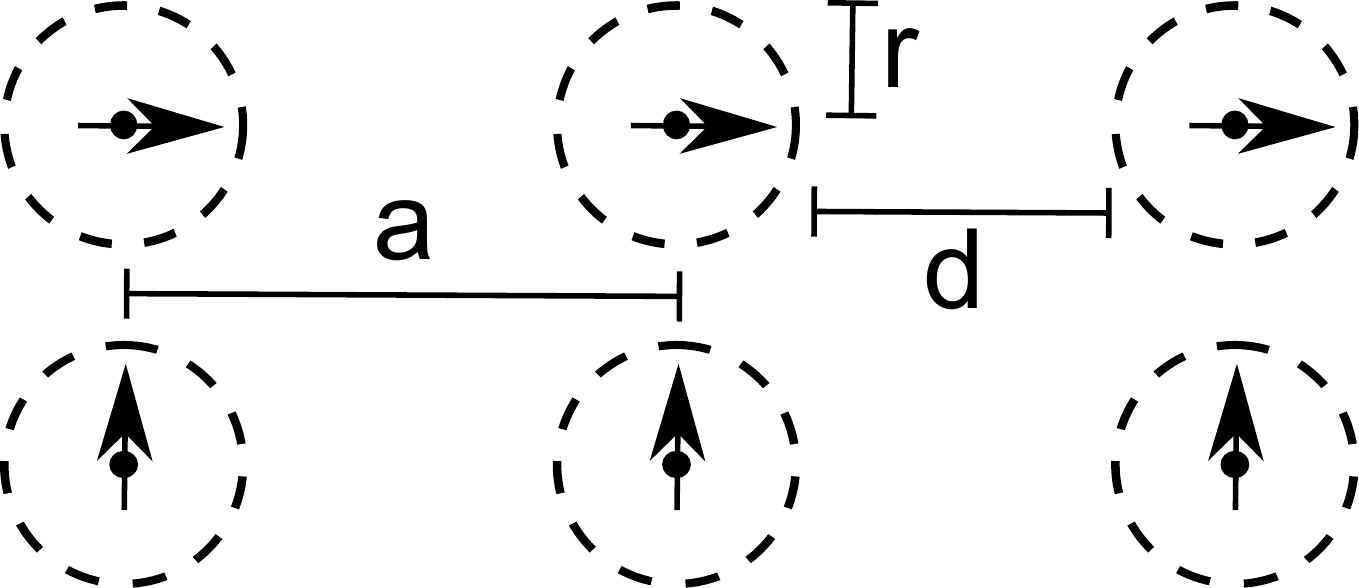}
  \end{centering}
  \caption{Chain of magnetic spheres of radius $r$, spaced at a distance
    $d$, approximated
    by point dipoles spaced by a distance $a=r+d$, magnetized along the
    longitudinal axis of the chain (top) or perpendicular to its
    longitudinal axis (bottom).}
  \label{fig:dipoles}
\end{figure}

The magnetosome does not consist of point dipoles but should be
approximated by spheres with
radius $r$, spaced at a distance of $a$ from each other
(figure~\ref{fig:dipoles}). We can modify the Jacob
and Bean model by introducing the volume of a single sphere $V$ and
the magnetization $M_s$ of the magnetite crystal (\SI{4.8e5}{A/m}~\cite{Witt2005}),

\begin{eqnarray}
\label{eq:25}
  \Gamma_\text{max}=\frac{1}{2}\mu_0M_s^2nV\Delta N\\
    \Delta N=2K_n\left(\frac{r}{a}\right)^3,
\end{eqnarray}

\noindent as a correction to equation~\ref{eq:22}.  This correction is based on
the fact that the field of a uniformly magnetized sphere is identical
to a dipole field~\cite{Griffiths1999book} outside the sphere, and the
average of the magnetic field over a sphere not containing currents is
identical to the field at the center of that
sphere~\cite{Hu2000,Hu2008}.

\begin{figure}
  \begin{centering}
    \includegraphics[width=\widefigurewidth]
    {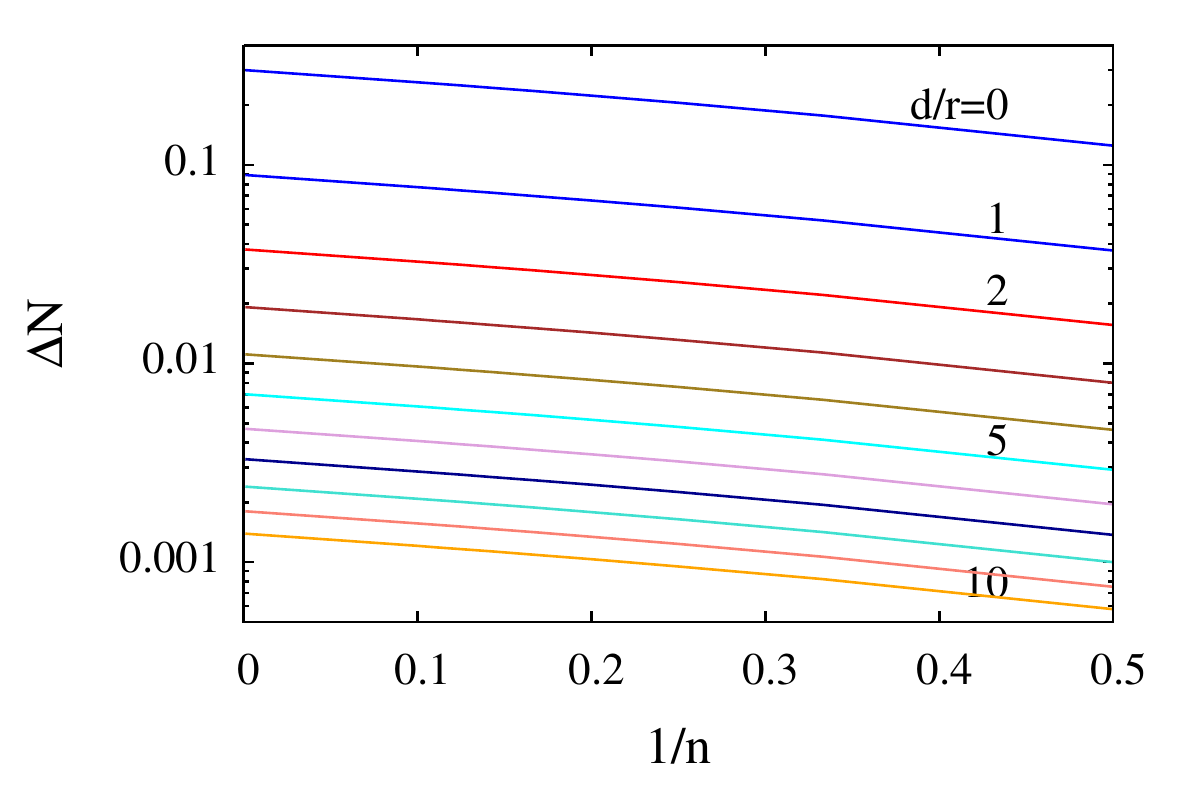}
  \end{centering}
  \caption{Difference in demagnetization factors of a chain of spheres
    as function of number of spheres $n$ for varying spacing between
    the spheres $d/r$.}
  \label{fig:Comparison_cal}
\end{figure}

For an infinitely long chain of touching spheres, $d$=0 and
$n\rightarrow\infty$, the difference in demagnetization factors
($\Delta N$) approaches $0.3$ (Figure~\ref{fig:Comparison_cal}). Approximating the chain by a
long cylinder ($\Delta N$=0.5)~\cite{Hanzlik1996, Erglis2007}
therefore overestimates the maximum torque by 40\%. Simply taking the total
magnetic moment to calculate the torque, as if $\Delta N$=1,
would overestimate it by a factor of three.

\subsubsection{Low field approximation}

For low values of the field ($h\ll1$), equation~(\ref{eq:22}) can be approximated by

\begin{align}
  \label{eq:linearapprox}
  \Gamma\approx\Gamma_\text{max}2h=\mu_0M_\text{s}nV H=m B,
\end{align}

\noindent where $m$ [\si{Am^2}] is the total magnetic moment of the magnetosome
chain and assuming the permeability of the medium to be equal to
vacuum. This approximation is commonly used in the field of MTB
studies. Based on the theory presented here, it is now possible to estimate up to
which field value this is approximation is valid.

\begin{figure}
  \begin{centering}
    \includegraphics[width=\widefigurewidth]{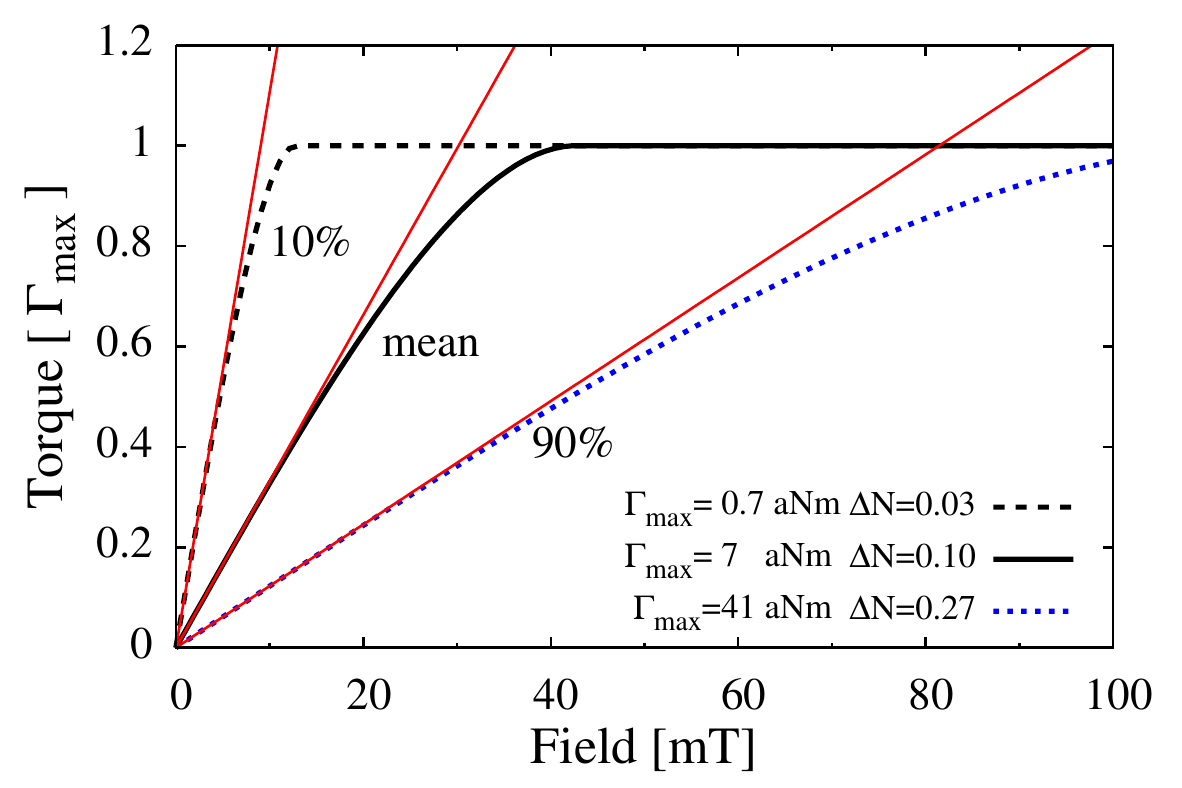}
  \end{centering}
  \caption{Magnetic torque on magnetotactic bacteria, normalized to the
    maximum torque, as a function of applied field for the average of
    the population, as well as the \SI{10}{\percent} and \SI{90}{\percent} cut-off (see
    table~\ref{tab:MTB_torque}). The red solid asymptotes show the
    linear approximation for $\Gamma=mB$.}
\label{fig:MTBTorqueVersusField}
\end{figure}

The normalization to the reduced field $h$ is solely dependent on the
magnetization and demagnetization factors of the chain.  Based on the
values for magnetosome morphology (table~\ref{tab:MTB_torque}), we can
estimate the field dependence of the torque.
Figure~\ref{fig:MTBTorqueVersusField} shows the torque as a function
of the field for the range of values tabulated, normalized to the maximum
torque.  Also shown is the approximation for the case when the
magnetization remains aligned with the easy axes. For
\emph{Magnetospirillum Gryphiswaldense}, the linear range is valid up
to fields of about \SI{10}{mT}  for \SI{90}{\percent} of the population.

\subsubsection{Drag torque}

Magnetotactic bacteria are very small, and rotate at a few
revolutions per second only. Inertial forces therefore do not play a
significant role. The ratio between the viscous and inertial forces is
characterized by the Reynolds number $Re$, which for rotation at an angular
velocity of $\omega$ [rad/s] is

\begin{equation}
\label{eq:Reynoldsnumber}
Re=\frac{L^2\rho\omega}{4\eta},
\end{equation}

\noindent where $L$ is the characteristic length (in our case, the length of the
bacterium), $\rho$ the density, and $\eta$ the dynamic viscosity
of the liquid (for water, these are, respectively, \SI{e3}{kg\per \cubic \meter}, and \SI{1}{mPas}). Experiments by Dennis
\emph{et al.}~\cite{Dennis1980} show that a Stokes flow approximation for
the drag torque is accurate up to $Re$=$10$. In experiments with
bacteria, the Reynolds number is on the order of \num{e-3} and the Stokes
flow approximation is certainly allowed. The drag torque is therefore
simply given by

\begin{equation}
\label{eq:rotationaldrag}
\Gamma_{\textrm{D}}=f_{\text{b}}\omega,
\end{equation}

The rotational drag coefficient of the bacterium, $f_\text{b}$, needs to be estimated for the type
of MTB studied. In a first approximation, one could consider the MTB to
be a rod of length $L$ and diameter $W$. Unfortunately, there is no
simple expression for the rotational drag of a
cylinder. Dote~\cite{Dote1983} gives a numerical estimate of the
rotational drag of a cylinder with spherical caps
(spherocylinder). Fortunately, for typical MSR-1 dimensions, it can be
shown that a prolate spheroid of equal length and diameter has a
rotational drag coefficient that is within \SI{10}{\percent} of that
value. To a first approximation, one can therefore assume the rotational drag of an
MSR-1 to be given by~\cite{Berg1993}

\begin{equation}
  \label{eq:dragcoef}
  f_{\text{e}}=\frac{\pi\eta L^{3}}{3\ln\left(\frac{2L}{W}\right)-\frac{3}{2}}.
\end{equation}

However, the MSR-1 has a spiral shape, so the actual drag will be
higher. Rather than resorting to complex finite element simulations,
we chose to empirically determine the rotational drag torque by
macroscopic experiments with 3D printed bacteria models in a highly
viscous medium (Section~\ref{sec:setup}.). We introduce a bacteria shape correction
factor $\alpha_\text{bs}$ to the spheroid approximation, which is
independent of the ratio $L/W$ over the range of typical values for
MSR-1 and has a value of about \num{1.65}. The corrected rotational drag
coeffient for the bacteria then becomes

\begin{equation}
  \label{eq:dragcoefMTB}
  f_{\textrm{b}}=\alpha_\text{bs}f_{\textrm{e}}.
\end{equation}

\subsubsection{Diameter and duration of the U-turn}

At the steady-state rate, the magnetic torque is balanced by the
rotational drag torque, leading to a rate of rotation of

\begin{equation}
\label{eq:maxrotfreq}
\omega=\frac{\Gamma}{f_\text{b}}\approx\frac{mB\sin\phi(t)}{f_\text{b}}.
\end{equation}

The approximation is for low field values (see
figure~\ref{fig:MTBTorqueVersusField}), in which case $\phi$ is the
angle between the applied field and the long axis of the bacteria (magnetosome).

The maximum rate of rotation, $mB/f_\text{b}$, is obtained when the field is
perpendicular to the long axis of the bacteria. Suppose that we construct a
control loop to realize this condition over the entire period of a
U-turn. Then the minimum diameter and duration of this loop
would be

\begin{align}
  \label{eq:D_min}
  D_{\text{min}}&=\frac{2f_\text{b}v}{mB}\\
  T_{\text{min}}&=\frac{\pi f_\text{b}}{mB},
\end{align}

\noindent where $D_\text{min}$ is the minimum size of a U-turn's diameter and $T_\text{min}$ is the minimum time of a U-turn. On the other hand, if we reverse the field instantaneously, the
torque will vary over the trajectory of the U-turn. Compared to the
situation above, the diameter of the U-turn increases by a factor of
$\nicefrac{\pi}{2}$:

\begin{equation}
  \label{eq:D_Esquivel}
  D=\frac{\pi f_\text{b}v}{mB}.
\end{equation}

The diameter of the U-turn increases with the velocity of the
bacterium. To obtain a description that only depends on the dimensions
of the bacteria, we introduce a new parameter $v/D$ [\si{rad/s}],
which can be interpreted as an average rate of rotation. The relation
between the average rate of rotation and the magnetic field $B$ is

\begin{equation}
  \label{eq:gammadefinition}
  \frac{v}{D}=\gamma B,
\end{equation}

\noindent where the proportionality factor $\gamma$ [\si{rad/Ts}] can be linked to the bacterial magnetic moment $m$ and drag coefficient
$f_\text{b}$ [\si{Nms}],

\begin{equation}
  \gamma=\frac{m}{\pi f_\text{b}}.
\end{equation}

Note, however, that this expression is only valid in the low field
approximation.

The determination of the duration of the U-turn trajectory is
complicated by the fact that the magnetic torque starts and ends at
zero (at $\theta$=0 or $\pi$). In this theoretical situation, the
bacteria would never turn at all. Esquivel \emph{et
  al.}~\cite{Esquivel1986} solve this problem by assuming a
disturbance acting on the motion of the bacteria. This disturbance could be
due to Brownian motion, as used by Esquivel~\emph{et al.}, or due to
flagellar propulsion, as we use in the simulations in the following
section. Assuming an initial disturbing angle of $\theta_\text{i}$,
the duration $T$ [s] of the U-turn becomes

\begin{equation}
  \label{eq:T_Esquivel}
  T=\frac{2f_\text{b}}{mB}\ln\frac{2}{\theta_\text{i}}.
\end{equation}

%%% Local Variables:
%%% mode: latex
%%% TeX-master: "UTurn"
%%% End:

% subsectie Tijmen
\subsection{U-turn Trajectory Simulations}
\label{sec:trajectories}

To check the validity of the analytical approach, we performed
simulations. The MTB are approximated by rigid magnetic
dipoles with constant lateral velocity $v$ at an orientation of $\theta_x(t)$
and angular velocity of $\omega(t)$ (see figure \ref{fig:bacAngles}). They are subject to a magnetic field
with magnitude $B$ at an orientation of $\varphi_x(t)$, resulting in a magnetic
torque of $\Gamma(t)$. In contrast to the analytical model, it
is assumed that flagellar motion causes an additive sinusoidal
torque $\Gamma_\text{f}(t)$ with amplitude $A_\text{f}$ and angular
velocity $\omega_\text{f}$. These should be in balance with the drag
torque: $\Gamma_\text{D}=f_\text{b} \omega(t)$. The following set of
equations link the physical model to the coordinates $x(t)$, $y(t)$:

\begin{figure}
  \begin{centering}
        \includegraphics[width=\narrowfigurewidth]{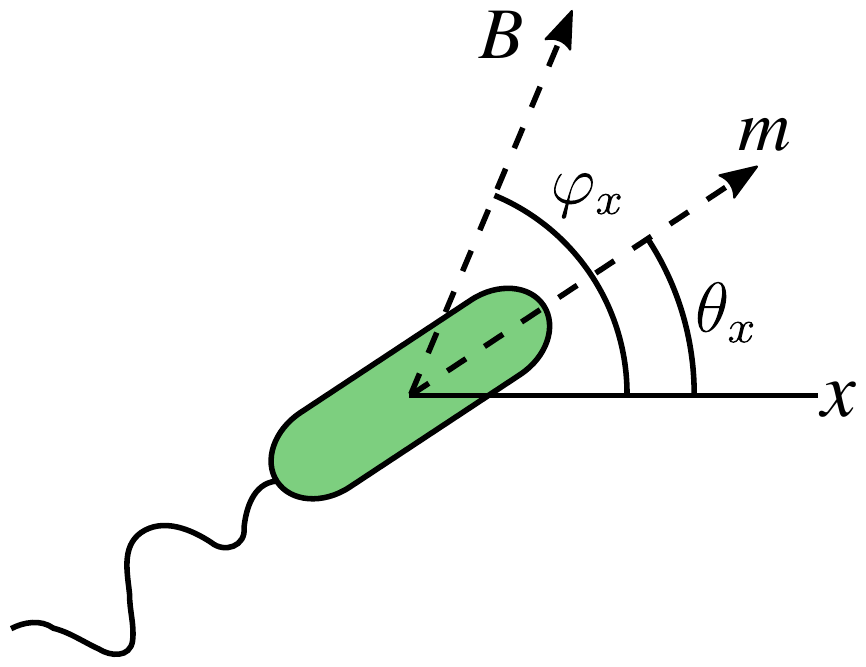}
  \end{centering}
  \caption{Bacterium at angle $\theta_x$ with magnetic field at angle $\varphi_x$.}
  \label{fig:bacAngles}
\end{figure}

\begin{align}
\label{eq:differential}
  x(t) &=& x(0) + \int_0^t v \cos(\theta_x(t)) \text{d}t\\
  y(t) &=& y(0) + \int_0^t v \sin(\theta_x(t)) \text{d}t\\
  \theta_x(t) &=& \theta_x(0) + \int_0^t \omega(t) \text{d}t \\
  \omega(t) &=& \frac{1}{f_\text{b}} \left( \Gamma_{\text{mag}}(t)
    +\Gamma_{\text{flag}}(t) \right) \\
  &=&\frac{m B}{f_\text{b}} \sin(\varphi_x(t) -
  \theta_x(t)) + \frac{A_\text{f}}{f_\text{b}} \sin(\omega_\text{f} t)
\end{align}

\begin{figure}
  \begin{centering}
        \includegraphics[width=\widefigurewidth]{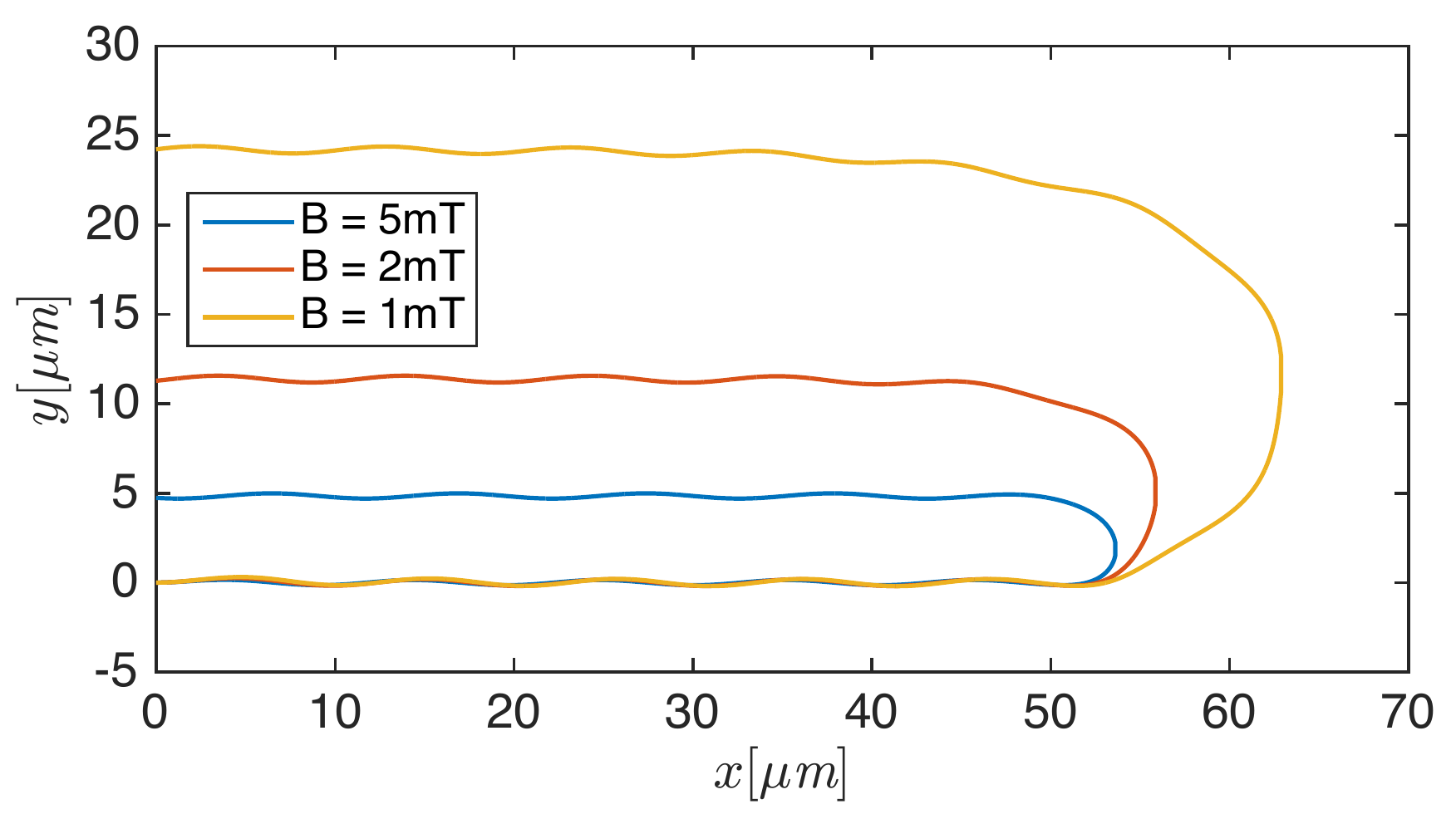}
  \end{centering}
  \caption{Simulated trajectories assuming flagellar torque and a
    non-instantaneously rotating field for several values of the
    magnetic field magnitude $B$. The time step of the simulation is \SI{10}{\micro\second}}
  \label{fig:trajectories}
\end{figure}

A linear, closed-form solution of the diameter of the trajectory of the U-turn
in the case of an instantaneous field reversal and no flagellar torque is 
given by equation \ref{eq:D_Esquivel}. This solution is not valid, however, in the case of slowly
rotating fields. The experimental magnetic field is considered to rotate according to a constant-acceleration model with a total rotation period of \SI{130}{ms} (see section~\ref{sec:setup}).
Simulations were carried out with time steps of \SI{10}{\micro\second}, which is comfortably fast and precise (decreasing this to \SI{1}{\micro\second} changes the results by approximately \SI{0.01}{\percent}). Figure \ref{fig:trajectories} shows several simulated trajectories subject to fields of various magnitudes, assuming nonzero flagellar torque and realistic MTB parameters.

\begin{figure}
  \begin{centering}
        \includegraphics[width=\widefigurewidth]{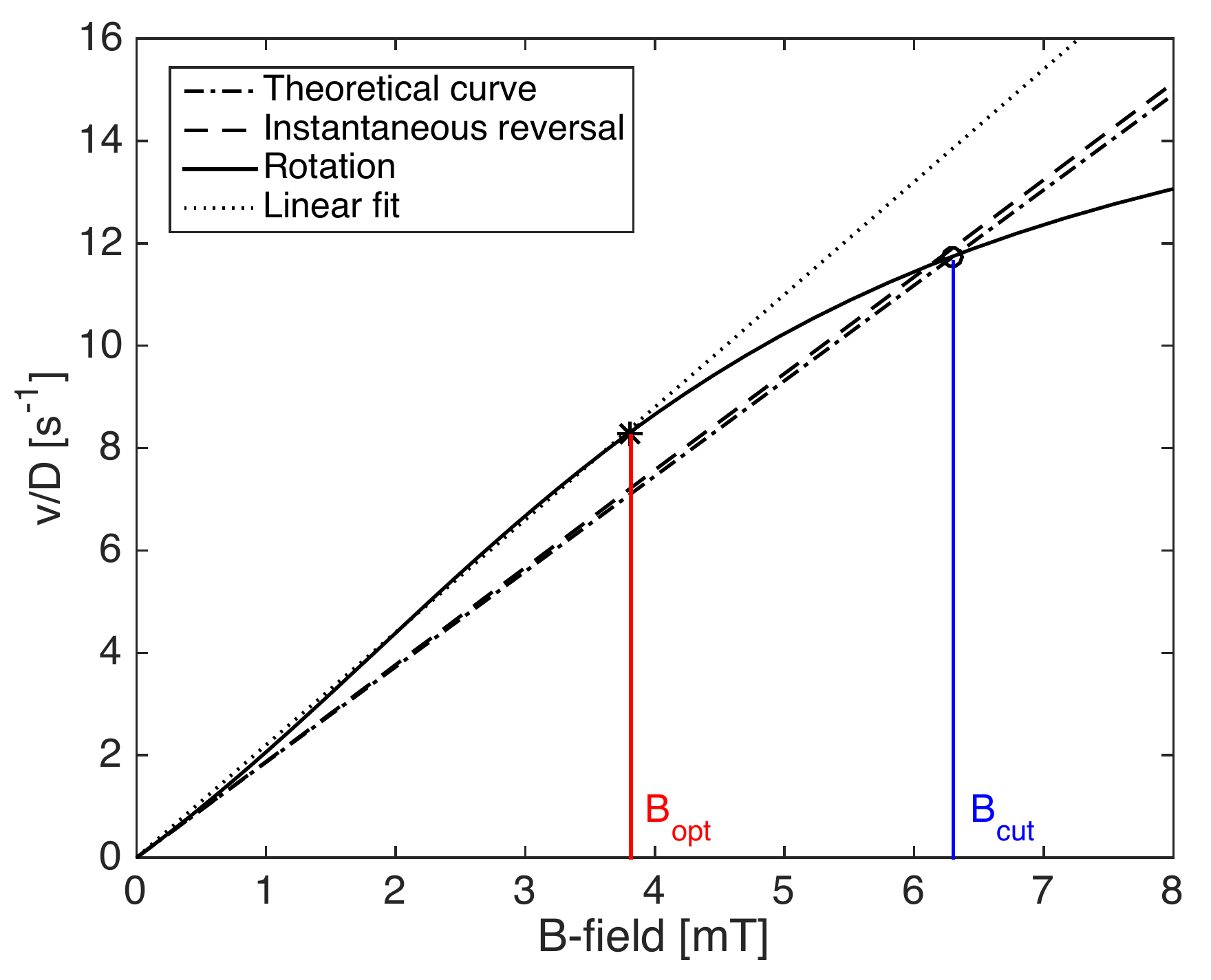}
  \end{centering}
  \caption{Simulated values of $v/D$ for different rotation speeds of the magnetic field,
    with (red) and without (blue) and flagellar torque,
    compared with the linear model proposed by Erglis \emph{et
      al.}~\cite{Erglis2007} (dotted line).}
  \label{fig:vR}
\end{figure}

\begin{figure}
  \begin{centering}
   \includegraphics[width=\widefigurewidth]{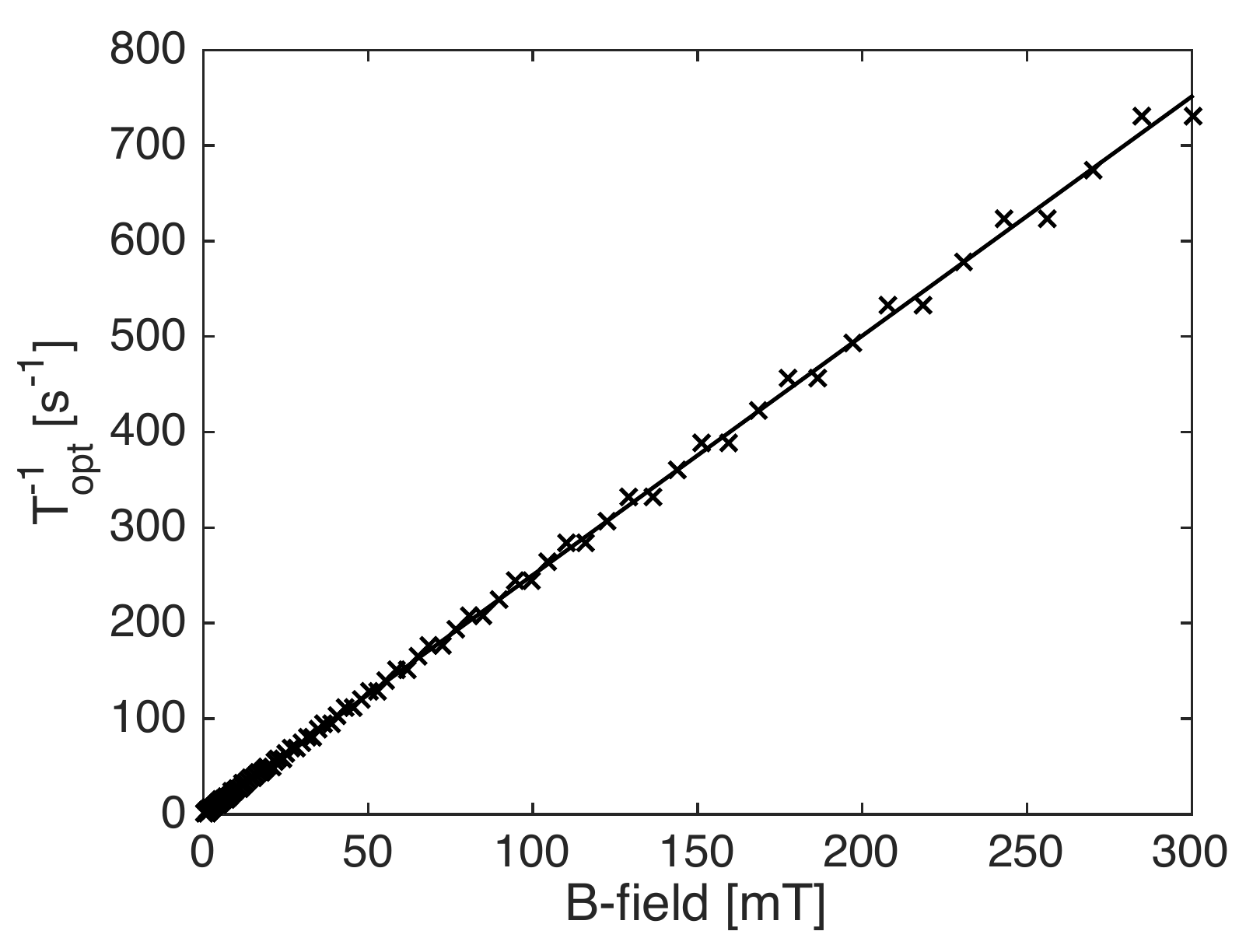}
  \end{centering}
  \caption{Simulated optimum reversal time of the magnetic field as a function of the field strength.}
  \label{fig:Tmag_opt}
\end{figure}

Figure \ref{fig:vR} shows the simulated $\nicefrac{v}{D}$ as a function of the
field magnitude. It can be seen that during
an instantaneous field reversal, the solution is nearly identical to the closed-form solution of equation \ref{eq:D_Esquivel}. The difference is caused by the influence of flagellar torque. Introducing a field
reversal time $T_{\text{mag}}$ of \SI{130}{ms} into a continuous-acceleration model significantly changes the profile, yielding a similar result for low fields, increasing at
moderate fields, and saturating to a maximum value of
\SI{16.6}{s^{-1}}. $B_{\text{opt}}$ is defined as the field magnitude
at which $\nicefrac{v}{D}$ has the largest difference from the
theoretical curve. Figure \ref{fig:Tmag_opt} shows, from simulations,
that the optimal reversal time is inversely proportional to the
magnetic field strength. For fields below $B_{\text{opt}}$,
$\nicefrac{v}{D}$ can be considered linear with a maximum nonlinearity
error of \SI{2}{\percent}, independently of $T_{\text{mag}}$.

%%% Local Variables:
%%% mode: latex
%%% TeX-master: "UTurn"
%%% End:

%%% Local Variables:
%%% mode: latex
%%% TeX-master: "UTurn"
%%% End:

\section{Experimental}
\label{sec:Experimental}

\subsection{Magnetotactic bacteria cultivation}

A culture of \emph{Magnetospirillum Gryphiswaldense} was used for the magnetic moment study. The cultures were inoculated in MSGM medium ATCC 1653 according to with an oxygen concentration of \SIrange{1}{5}{\percent}. The bacteria were cultivated at \SI{21}{\celsius} for \SIrange{2}{5}{days} for optimal chain growth~\cite{Katzmann2013}. The sampling was done using a magnetic ``racetrack'' separation, as described in ~\cite{Wolfe1987}.

\subsection{Dynamic viscosity of growth medium}
The kinematic viscosity of the freshly prepared growth medium was
determined with an Ubbelohde viscometer with a capillary diameter
of \SI{0.63(1)}{mm} (Si Analytics 50110). The viscometer was
calibrated with deionized water, assuming it has a kinematic viscosity of
\SI{0.98(1)}{mm^2/s} at \SI{21.0(5)}{\celsius}. At that temperature, the
growth medium has a kinematic viscosity of \SI{0.994(17)}{mm^2/s}. The
density of the growth medium was \SI{1.009(2)}{g/cm^3}, measured by
weighing \SI{1}{ml} of it on a balance. The dynamic viscosity of the
growth medium is therefore \SI{1.004(19)}{mPas}, which is, within
measurement error, identical to water (\SI{1.002}{mPas}).

\subsection{Microfluidic Chips}
\begin{figure}
  \begin{center}
    \includegraphics[width=\widefigurewidth]{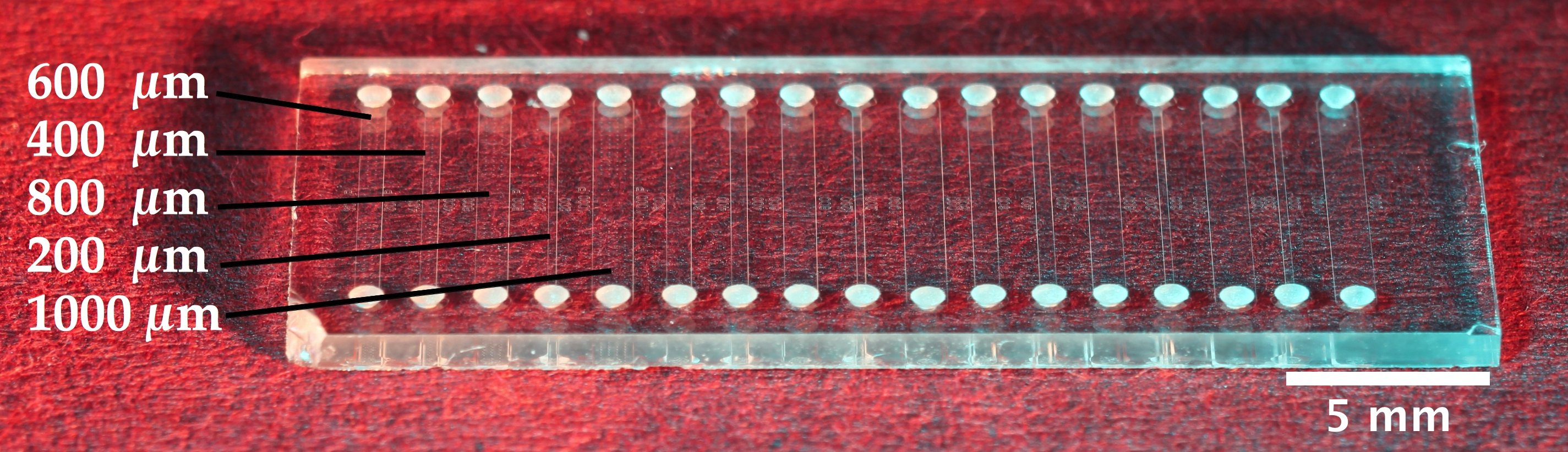}
    \caption{Top: A \SI{5}{\um} deep microfluidic chip with various channel
      widths of 200, 400, 600, 800 and 1000 \si{\um}.}
    \label{fig:chip}
  \end{center}
\end{figure}

Microfluidic chips with a channel depth of \SI{5}{\um} were constructed by lithography, HF etching in glass and subsequent thermal bonding. The fabrication process is identical to the one described in~\cite{Park2015}.  Figure~\ref{fig:chip} shows the resulting structures, consisting of straight channels with inlets on both sides. By means of these shallow channels, the MTBs are kept within the field of focus during microscopic observation, so as to prevent out-of-plane focus while tracking. The channel width was \SI{200}{\um} or more, so that the area over which U-turns could be observed was only limited by the field of view of the microscope.

\subsection{Magnetic Manipulation Setup}
\label{sec:setup}

\begin{figure}
  \begin{centering}
        \includegraphics[width=\widefigurewidth]{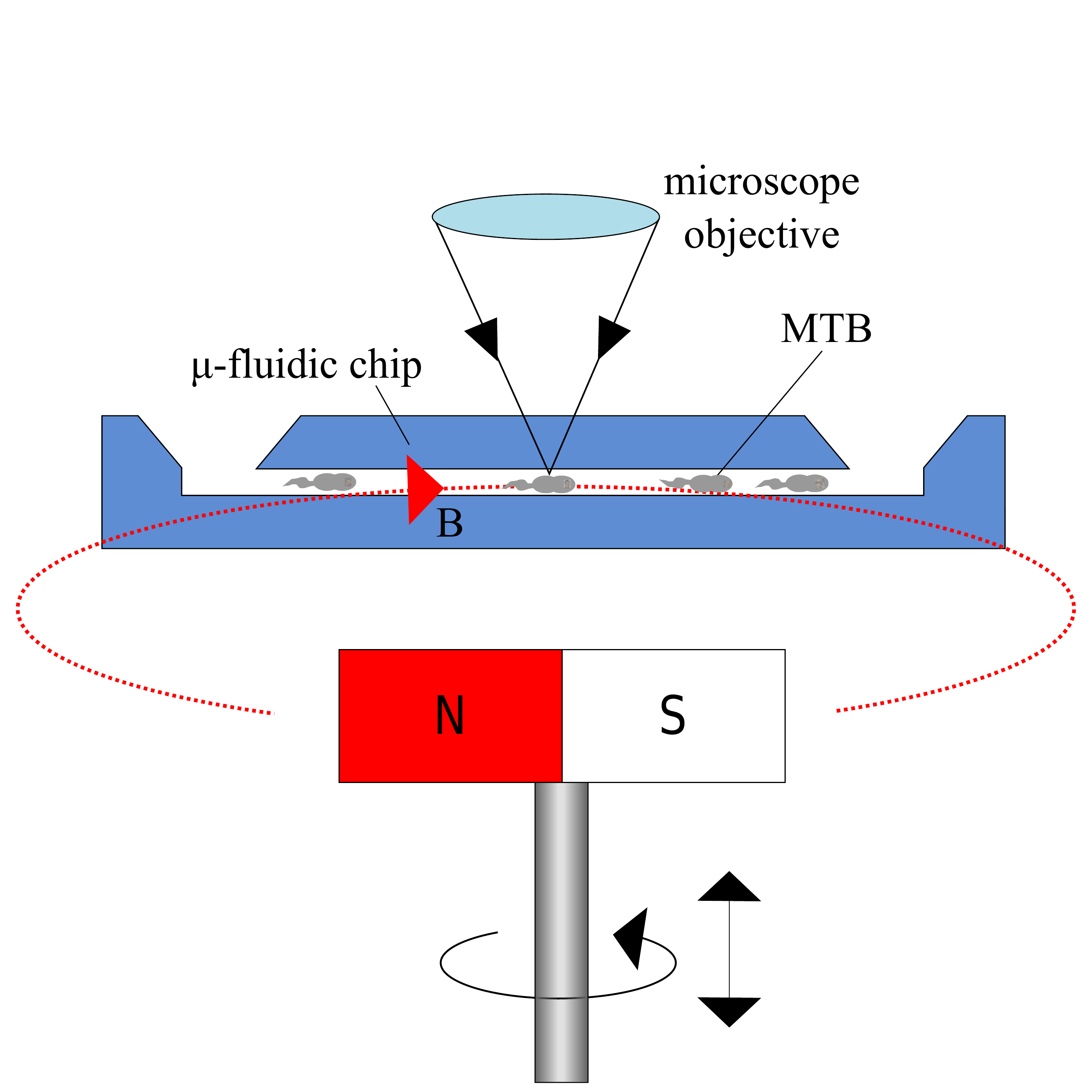}
  \end{centering}
  \caption{The setup used to measure the MTB U-turns. (a) Reflective
    microscope, (b) microfluidic chip and (c) a
    permanent magnet mounted on (d) a stepper motor.}
  \label{fig:setupmag}
\end{figure}

A schematic of the full setup, excluding the computer used for the acquisition of the
images, is shown in figure~\ref{fig:setupmag}. A permanent NdFeB magnet
(\SI{5x5x10}{mm}, grade N42) is mounted on a stepper motor (Silverpak
17CE, Lin Engineering) below the microfluidic
chip. The direction of the field can be adjusted with a precision of
\num{51200} steps for a full
rotation, at a rotation time of \SI{130}{ms} with a constant
acceleration of \SI{745}{rad s^{-2}}. The field strength is adjusted
using a labjack, with a positioning accuracy of \SI{0.5}{mm}.

The data acquisition was done by a Flea3 digital camera (\num{1328}$\times$\num{1048} at
\SI{100}{fps}, FL3-U3-13S2M-CS, Point Grey) mounted on a Zeiss
Axiotron 2 microscope with a \num{20}$\times$ objective.

During the experiments, a group of MTB was observed while periodically (every two seconds) rotating
the magnetic field. This was recorded for field magnitudes
ranging from \SIrange{1}{12}{mT}. Offline image processing
techniques were used to track the bacteria and subtract their velocity
and U-turn diameter.

\begin{figure}
  \begin{centering}
        \includegraphics[width=\figurewidth]{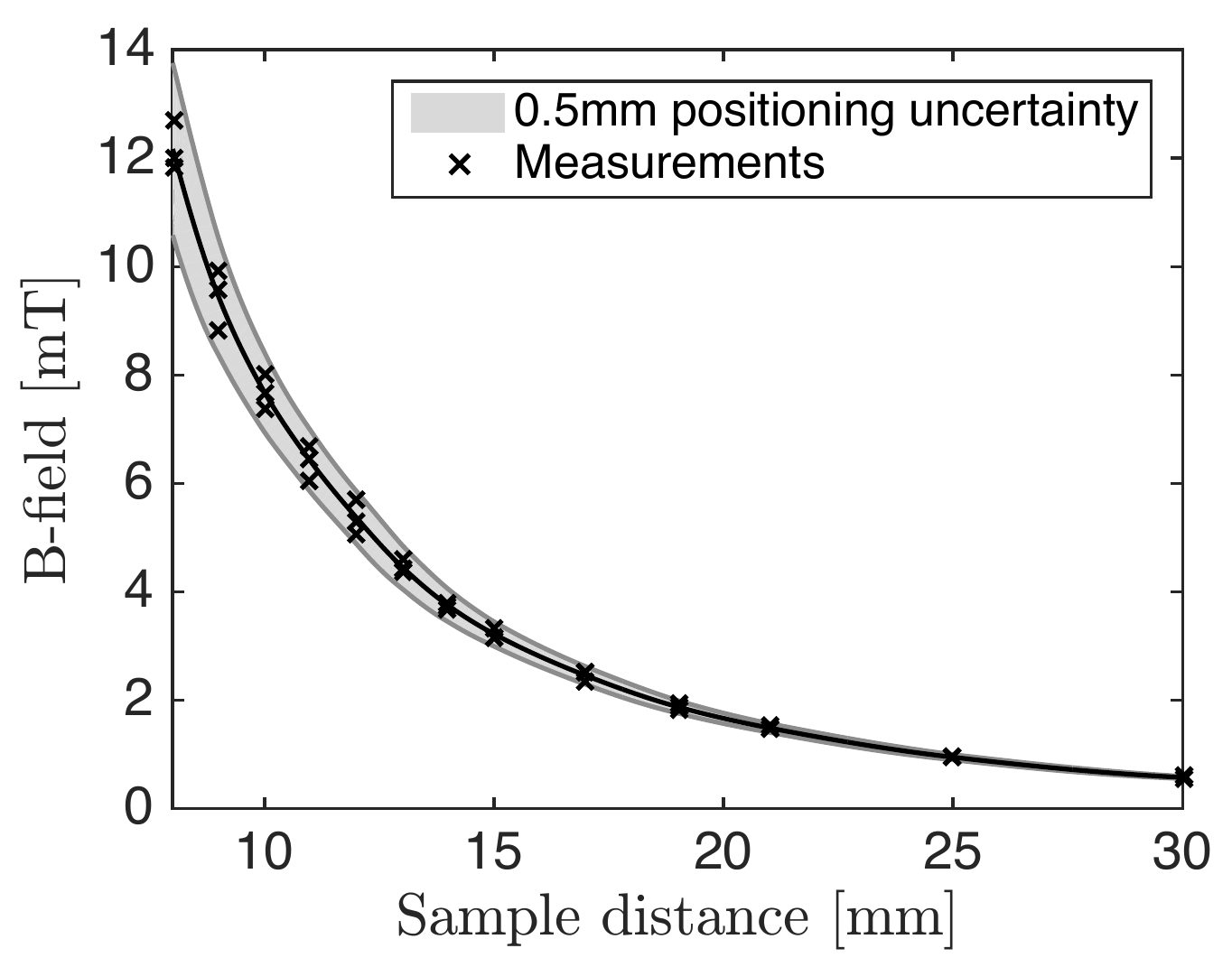}
  \end{centering}
  \caption{Magnetic field strength as a function of distance of the
    magnet to the microfluidic chip.}
  \label{fig:mag_positioning}
\end{figure}

Knowing the error in our measurements of the magnetic field is fundamental to determining the responsiveness of the MTB. Therefore we measured the magnetic fields at specific heights using a Hall meter (Metrolab THM1176). The results can be seen in figure~\ref{fig:mag_positioning}.

The placement of the tip of the Hall meter was at the location of the microfluidic chip, assuming the field strength inside the chip's chamber equals that at the tip. It should be noted that the center of the magnet was aligned with the center of rotation of the motor, therefore the measurements were only done with a stationary magnet on top of an inactive motor. Errors in the estimation of the magnetic field strength due to misalignment of the magnetic center from our measurements therefore cannot be excluded.

\begin{figure}
  \begin{centering}
        \includegraphics[width=\figurewidth]{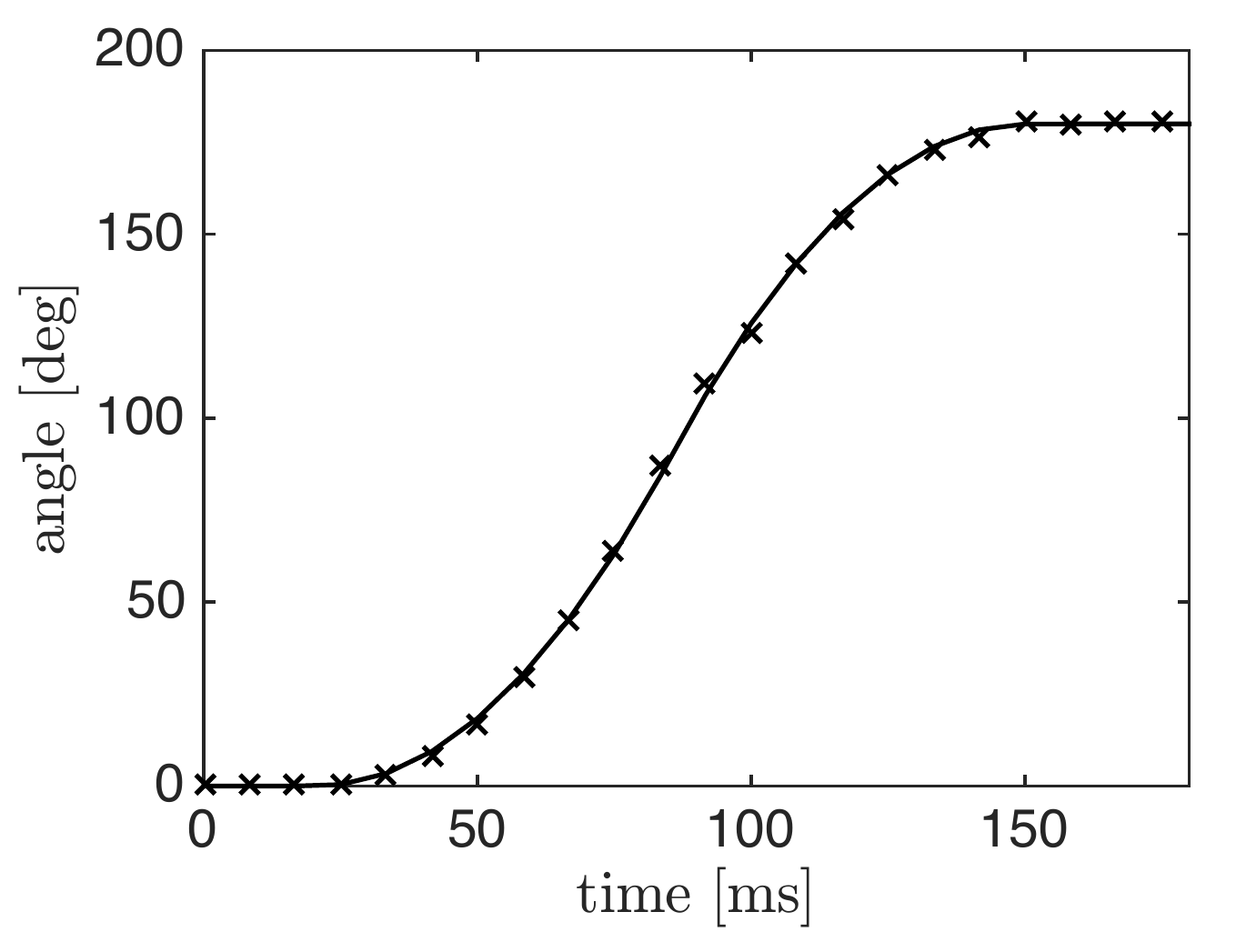}
  \end{centering}
  \caption{The measured angle of the motorized magnet accurately
  fits a constant-acceleration model with a total rotation period of \SI{130}{ms}.}
  \label{fig:motor_rotation}
\end{figure}

The rotation profile of the motorized magnet was investigated by
recording its motion by a digital camera at \SI{120}{fps} and
evaluating its time-dependent angle by manually drawing tangent
lines. Figure~\ref{fig:motor_rotation} shows that the profile accurately fits
a constant-acceleration model with an acceleration of
\SI{745}{rad s^{-2}}, resulting in a total rotation time of
\SI{130}{ms}. 

\subsection{Macroscopic Drag Setup}
\label{sec:dragsetup}

\begin{figure}
  \begin{center}
    \includegraphics[width=\widefigurewidth]{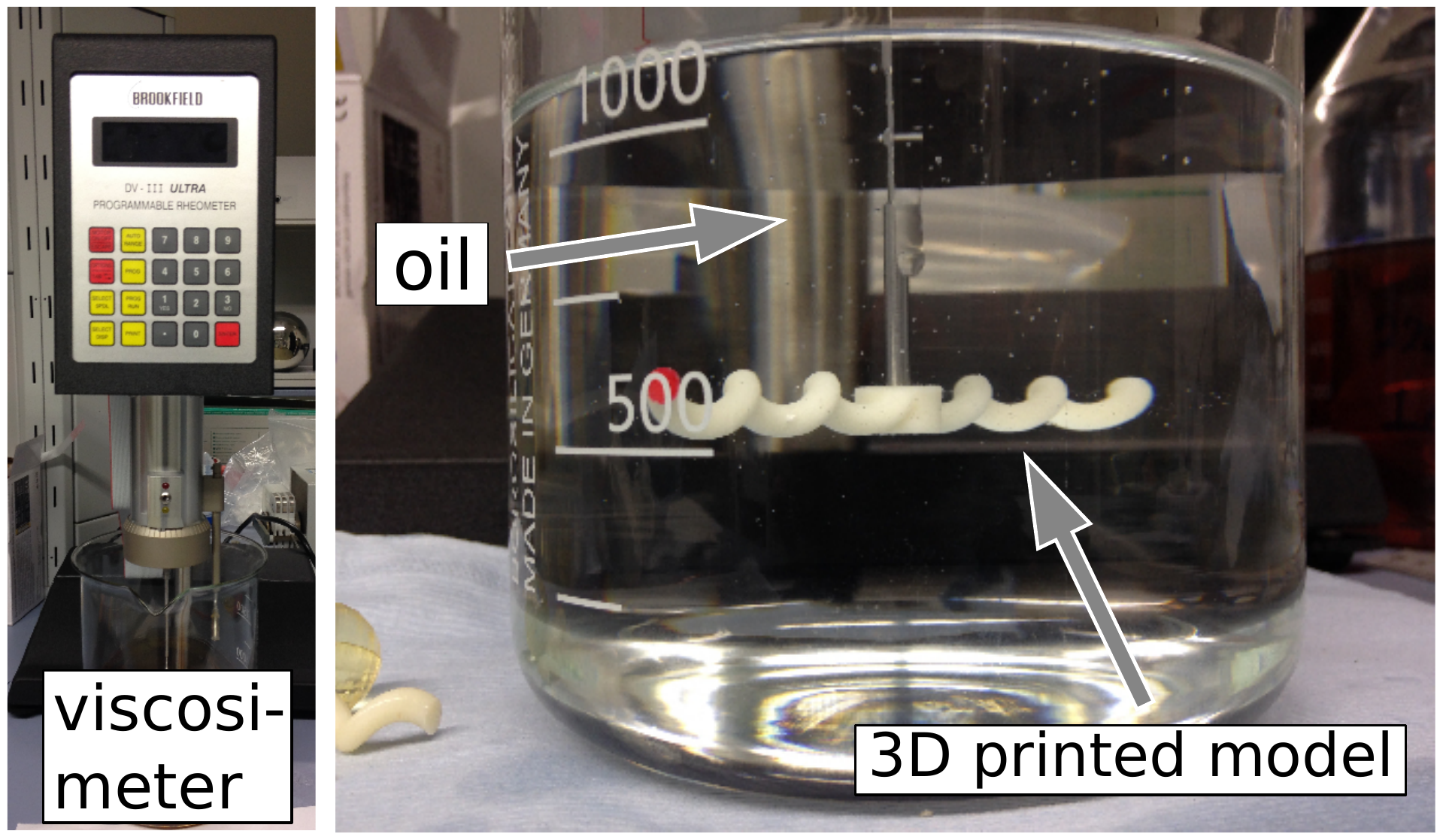}
    \caption{The viscometer setup used to measure the
      rotational drag of macroscopic spheroid and helical
      structures. 3D printed models were mounted on a shaft and
      rotated in a high viscosity silicone oil (\SI{5}{Pas}). A video of the
      experiment is available as additional material
      (DragMeasurements.mp4).}
    \label{fig:viscometer}
  \end{center}
\end{figure}

Macro-scale drag measurements were performed using a Brookfield DV-III
Ultra viscometer. During the experiment, we measured the torque
required to rotate different centimeter sized models of bacteria and
simple shapes in silicone oil (Figure~\ref{fig:viscometer}). In order
to keep the Reynolds number less than one, silicone oil of
\SI{5000}{mPas} (Calsil IP 5000 from Caldic, Belgium) was used as a
medium to generate enough drag. Furthermore, the parts were rotated at
speeds below \SI{30}{rpm}. The models were realized by 3D
printing. The designs can be found in the accompanying material.

\subsection{Image Processing}
\begin{figure}
  \begin{centering}
        \includegraphics[width=\widefigurewidth]{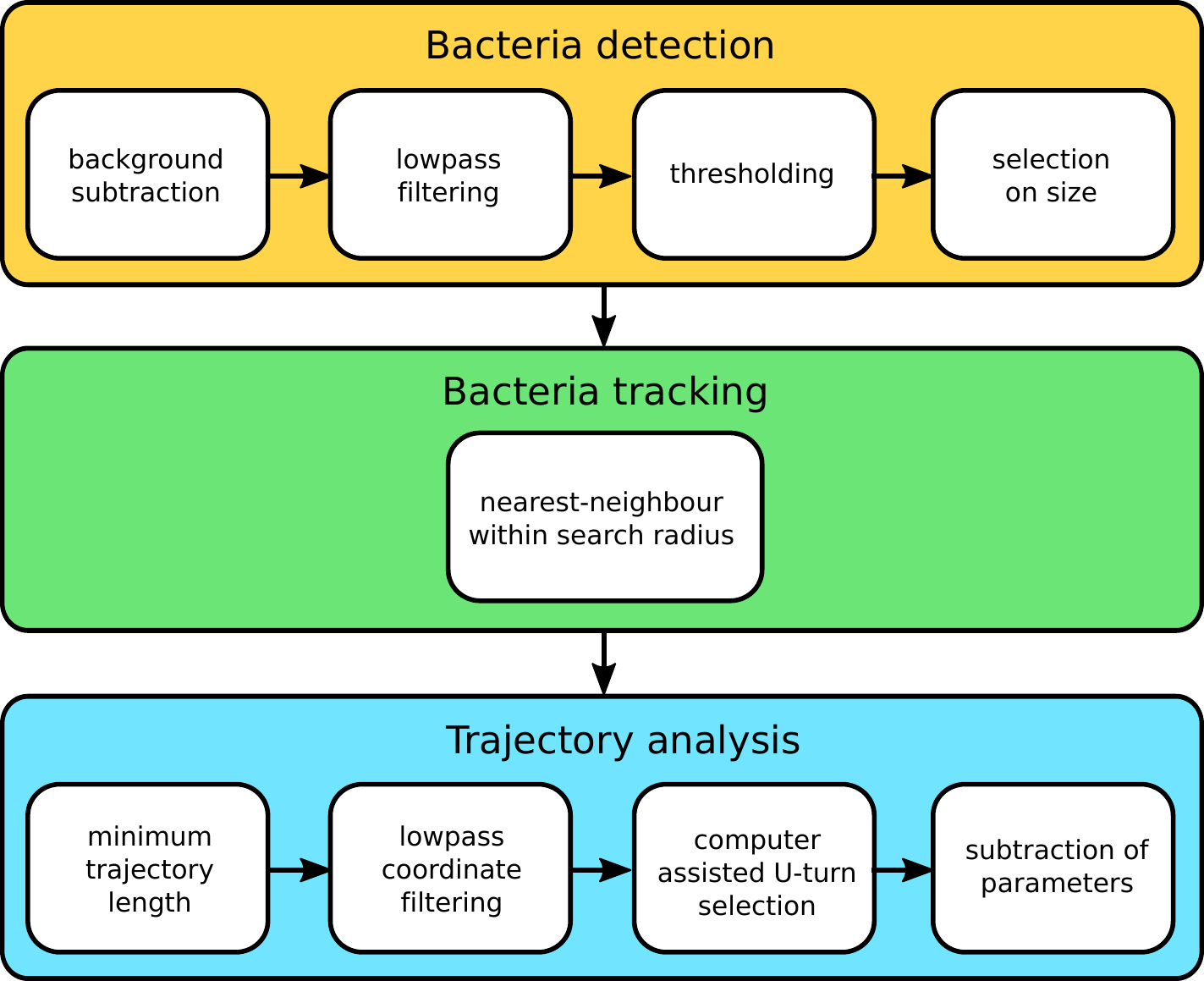}
  \end{centering}
  \caption{The process of bacteria detection, tracking, and subsequent analysis.}
  \label{fig:video_analysis}
\end{figure}

\begin{figure*}
  \begin{centering}
    \includegraphics[width=2\widefigurewidth]{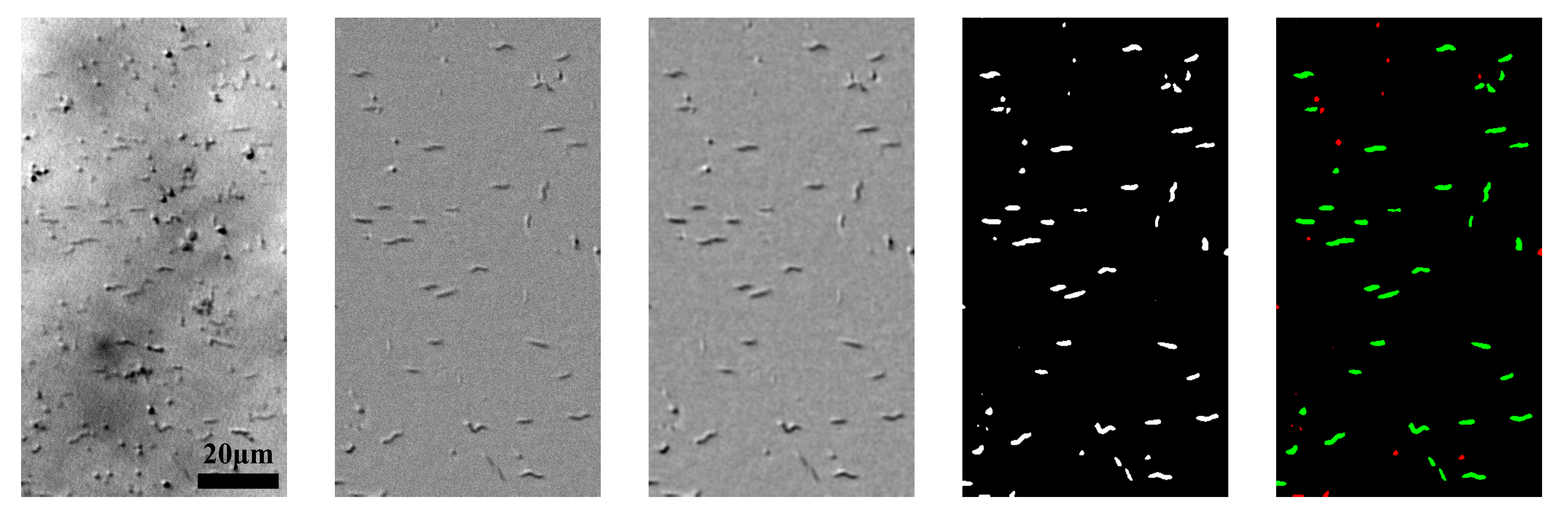}
  \end{centering}
  \caption{Pre-processing filter steps: (a) raw, (b) background subtraction,
    (c) low pass filtering, (d) thresholding resulting in a binary
    image, (e) size selectivity. A video is available as additional material ($MTB\_imageProc.mp4$)}
\label{fig:preproc}
\end{figure*}

The analysis of the data was done using in-house detection and tracking
scripts written in MATLAB\textsuperscript{\textregistered}. The process is illustrated in figure \ref{fig:video_analysis}. In the
detection step, static objects and non-uniform illumination artefacts are
removed by subtracting a background image constructed by
averaging 30 frames spread along the video. High-frequency noise is
reduced using a Gaussian lowpass filter. A binary image is then obtained using
a thresholding operation, followed by selection on a minimum and maximum
area size. The centers-of-mass of the remaining blobs are compared
in subsequent frames, and woven to trajectories based on a
nearest-neighbor search within a search radius \ref{fig:video_analysis}. A sequence of preprocessing steps can be seen in figure~\ref{fig:preproc}. The software used is available under additional material.

\begin{figure*}
    \centering
    \centering
        \includegraphics[width=\widefigurewidth]{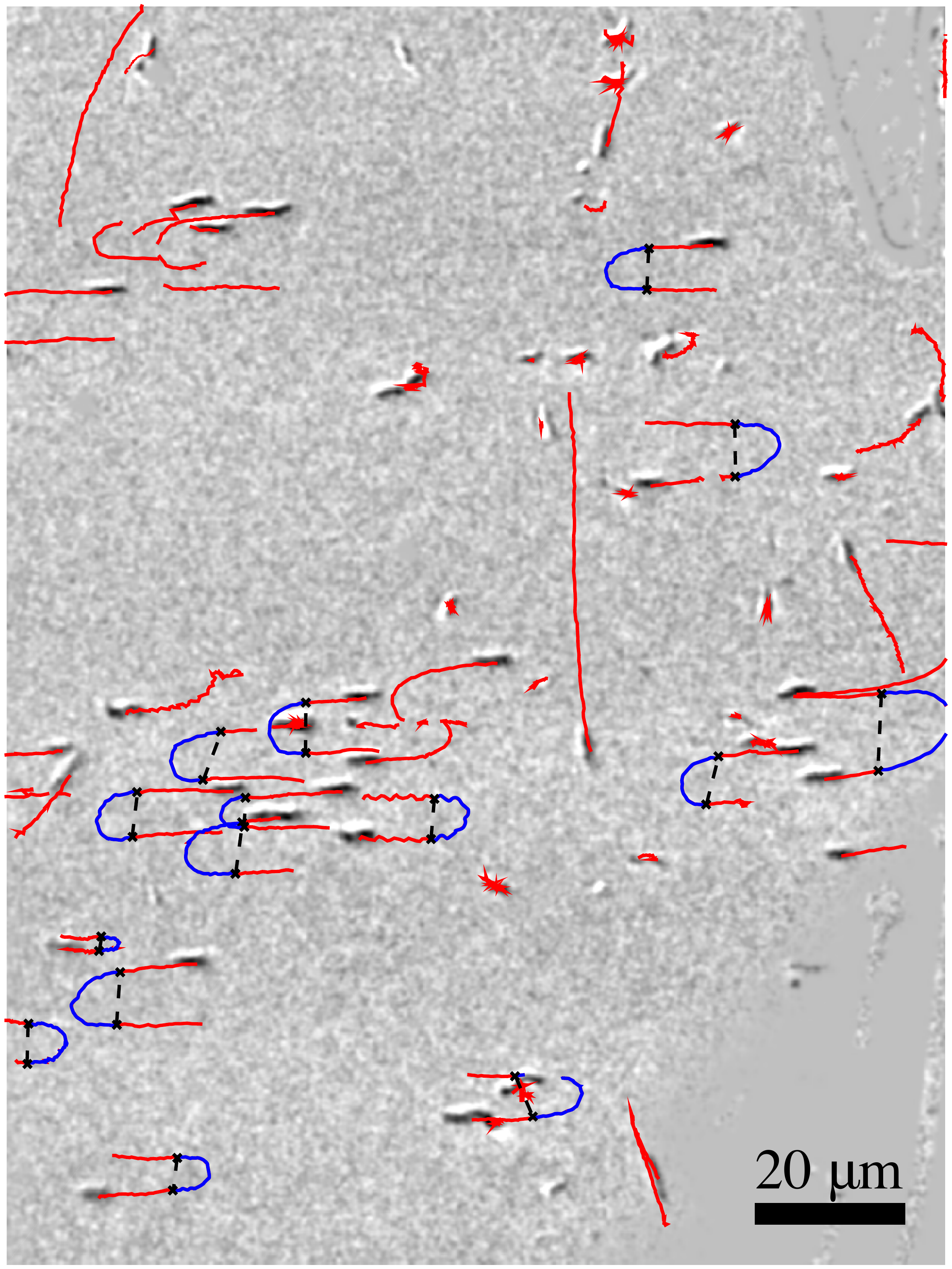}
        \includegraphics[width=\widefigurewidth]{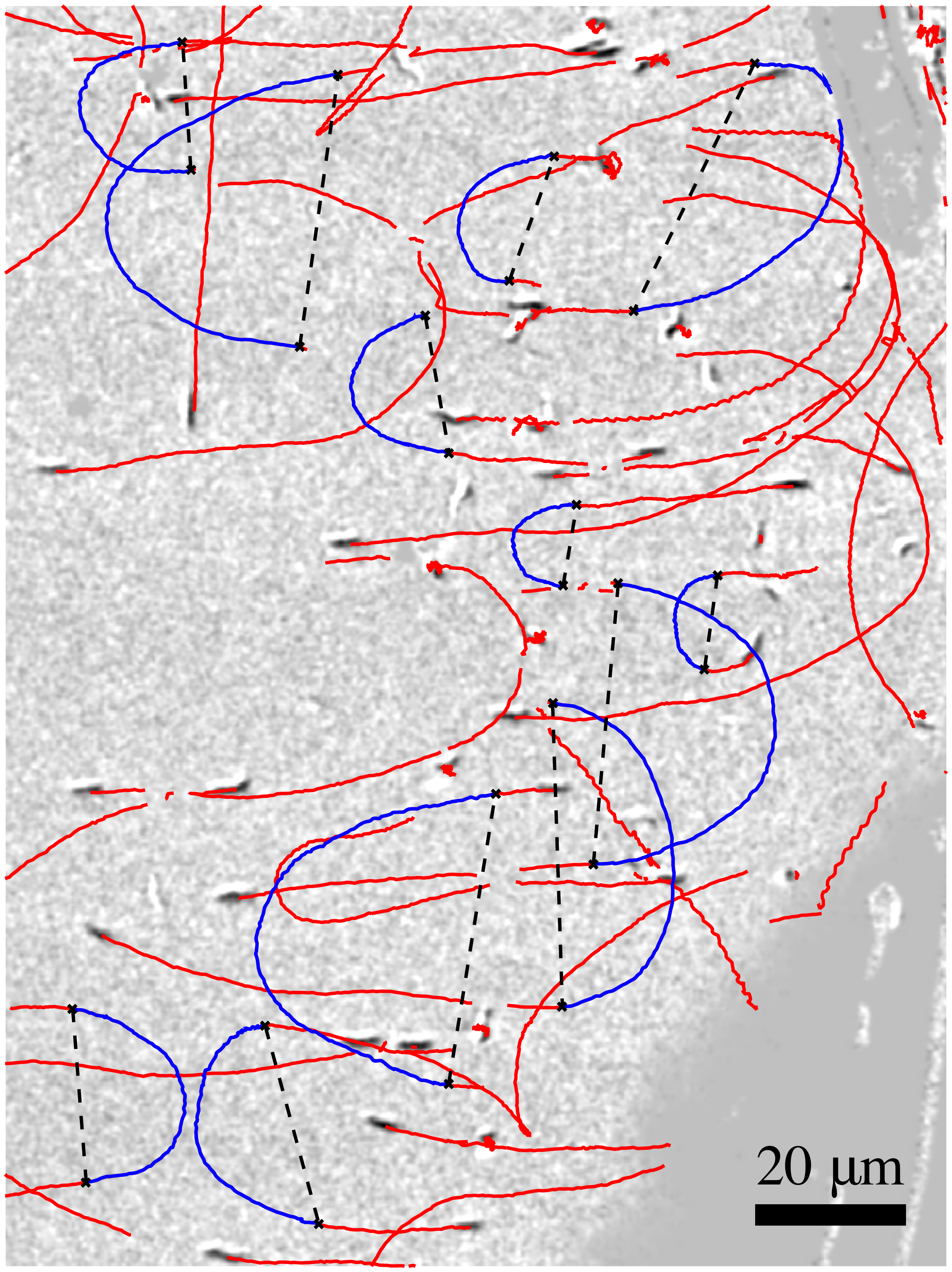}
        \caption{Trajectory during image post-processing at a magnetic
          field strength of \SI{12.2}{\milli\tesla} (left) and
          \SI{1.5}{\milli\tesla} (right). Selection procedure of
          analyzed U-turns, showing selected U-turns in blue and
          unanalyzed trajectories in red. The black dotted line
          connects two manually selected points of a given U-turn
          trajectory, from which the distance in the $y$-direction, or
          the U-turn diameter, is determined.}
      \label{fig:trajectoryselect}
\end{figure*}

Subsequently, the post-processing step involves the semi-automated selection of the MTB
trajectories of interest for the purpose of analysis. The U-turn parameters of interest analyzed are the velocity $v$, the diameter $D$ of the U-turn, and the time $t$. A typical result of the post-processing step can be seen in figure~\ref{fig:trajectoryselect}.

%%% Local Variables:
%%% mode: latex
%%% TeX-master: "UTurn"
%%% End:

\section{Results and discussions}
\label{sec:results}
The model developed in section~\ref{sec:theory} predicts the
trajectories of MTB under a changing magnetic field: in particular,
the average rate of rotation over a U-turn. To validate the model, the
essential model parameters are determined in
section~\ref{sec:estim-prop-fact}, after which the average rate of
rotation is measured and compared to theory in section~\ref{sec:trajectories-1}.

% Estimate of bacteria and magnetosome  size, leading to estimate of
% average rate of rotation.
\subsection{Estimate of model parameters}
\label{sec:estim-prop-fact}
The rate of rotation of an MTB under a rotating magnetic field
is determined by the ratio between the rotational drag torque and the magnetic
torque. Both will be discussed in the following, after which the
average rate of rotation will be estimated.

\subsubsection{Estimate of rotational drag torque}
\label{sec:dimens-bact-their}
To determine the rotational drag torque, the outer shape of the MTB
was measured by both optical microscopy and scanning electron
microscopy (SEM). The drag coefficient was estimated from
a macroscale drag viscosity measurement.

\paragraph{Outer dimensions of the bacteria }
The length $L$ of the bacteria is measured from the same optical
images as used for the trajectory analysis
(figures~\ref{fig:trajectoryselect}).  Scanning electron
microscopy (SEM) would in principle give higher precision per
bacterium, but due to the lower number of bacteria per image the
estimate of the average length and distribution would have a higher
error. Moreover, using the video footage ensures that the radius of
curvature and the length of the bacteria are measured on the same bacterium.

\begin{figure}
  \begin{centering}
    \includegraphics[width=\widefigurewidth]{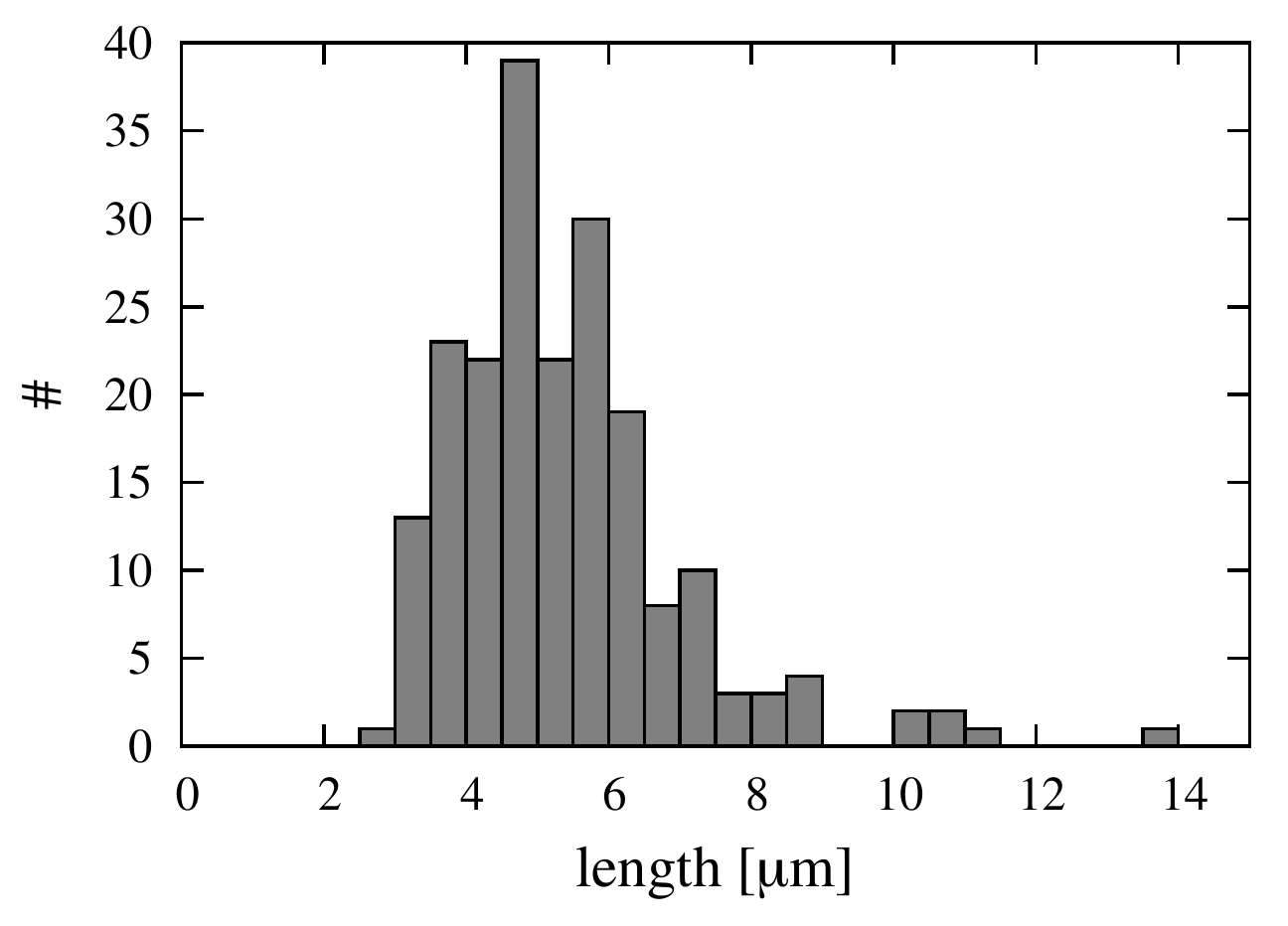}
  \end{centering}
  \caption{Number of magnetotactic bacteria (MTB) as a function of the length of the MTB as measured by optical microscopy.}
\label{fig:MTBlength}
\end{figure}

A typical MTB has a length of \SI{5.0(2)}{\um}. The length
distribution is shown in Figure~\ref{fig:MTBlength}. These values agree
with values reported in the literature~\cite{Schleifer1991,
  Bazylinski2004, Faivre2010}.

\begin{figure}
  \begin{centering}
    \includegraphics[width=\widefigurewidth]{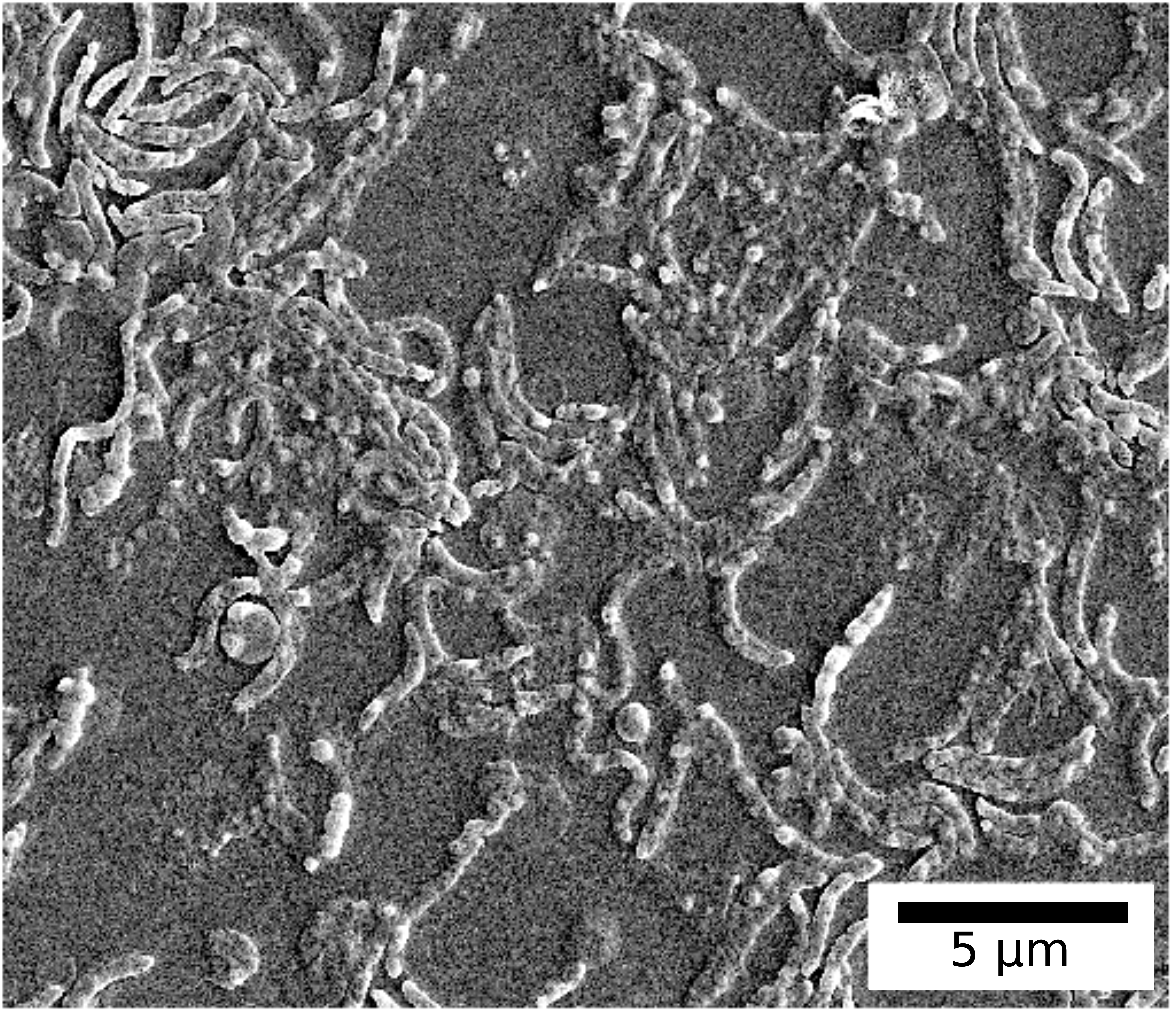}
  \end{centering}
  \caption{Scanning electron microscopy images of \emph{Magnetospirillum
      Gryphiswaldense}. Separated MTB were selected for width
    measurements.}
\label{fig:MTBSEM}
\end{figure}

The width $W$ of the bacteria is too small to be determined by optical
microscopy, and needs to be determined from SEM images, see figure~\ref{fig:MTBSEM}. A typical
bacterium has a width of \SI{240(6)}{nm}.  The main issue with SEM
images is whether a biological structure is still intact or perhaps
collapsed due to dehydration, which might cause overestimation of the
width. The latter might be as high as $\nicefrac{\pi}{2}$ if the
bacterial membrane has completely collapsed. Fortunately, the drag
coefficient scales much more strongly with the length than with the width
(equation~\ref{eq:dragcoef}). For a typical bacterium, the
overestimation of the width by using SEM leads at most to an
overestimation of the drag by \SI{18}{\percent}.

Table~\ref{tab:MTB_torque} lists the values of the outer dimensions
$L$ and $W$, including the measurement error and standard deviation
over the measured population.

\begin{figure}
  \begin{centering}
    \includegraphics[width=\widefigurewidth]
    {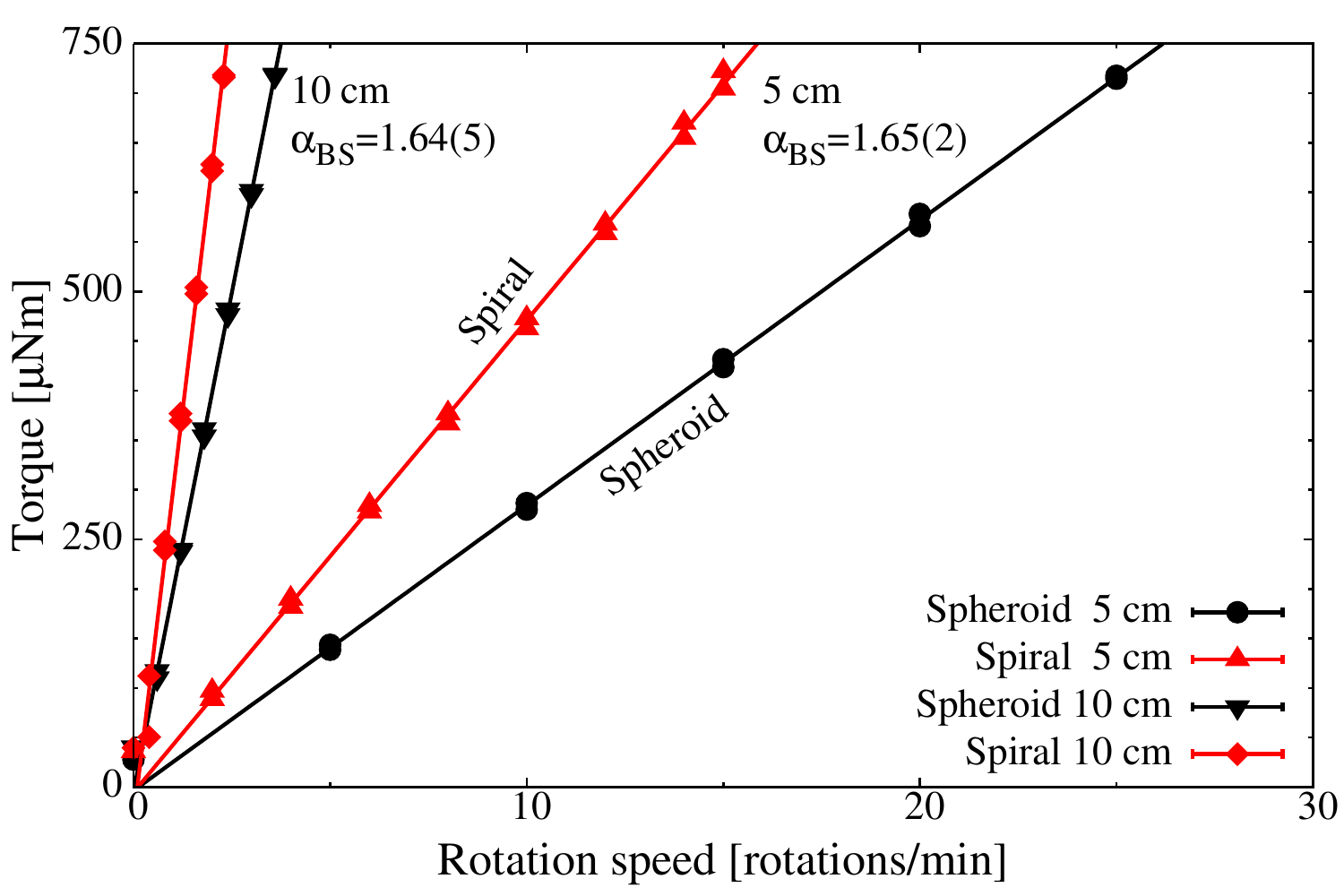}
  \end{centering}
  \caption{Rotational drag torque versus angular rotation speed of 3D
    printed prolate spheroids and MTB models of lengths \num{5} and \SI{10}{cm}. The
    curves are linear, indicating that the flow around the objects is
    laminar. Irrespective of the length, the spiral shaped MTB model has a
    drag that is \SI{1.64(5)}{} higher than a prolate spheroid of equal
    overall length and diameter.}
  \label{fig:MTBandSpheroids}
\end{figure}

\paragraph{Rotational Drag}
From the outer dimensions of the bacteria, the rotational drag torque can be
estimated. The bacterial shape correction factor,
equation~(\ref{eq:dragcoefMTB}), was determined by macro-scale
experiments with 3D printed models of an MTB in a viscosimeter using
high viscosity silicone oil (see
section~\ref{sec:dragsetup}). Figure~\ref{fig:MTBandSpheroids} shows
the measured torque as a function of the rotational speed for
prolate spheroids and spiral shaped 3D printed bodies of two different
lengths. The relation between the torque and the speed is linear, so we are
clearly in the laminar flow regime. This is in agreement with an
estimated Reynolds number of less than \SI{0.3}{} for this experiment
(equation~\ref{eq:Reynoldsnumber}). Independently
of the size, the spiral shaped MTB models have a drag coefficient that
is \SI{1.64(5)}{} times higher ($\alpha_\text{BS}$) than that of a spheroid
of equal overall length and diameter.

\begin{figure}
  \begin{centering}
    \includegraphics[width=\widefigurewidth]
    {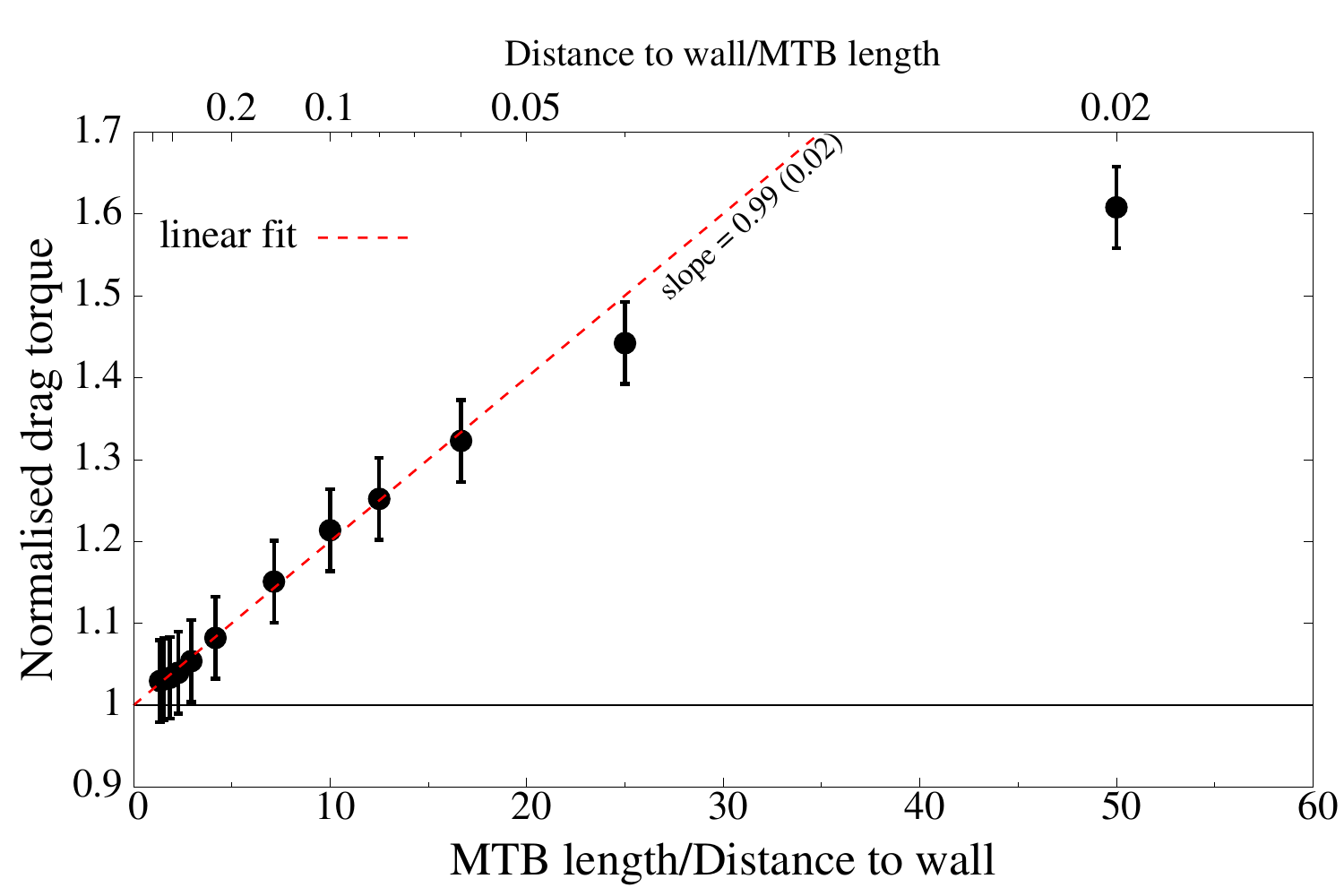}
  \end{centering}
  \caption{Increase in rotational drag as a function of the distance
    between the 3D printed spiral and the bottom of the container. The
    distance is  normalized to the length of the bacteria (\SI{5}{cm}). The
    torque is normalized to the extrapolated value for infinite
    distance (displayed as ``linear fit'').}
  \label{fig:walleffect}
\end{figure}

Using the same experimental configuration, we can obtain an estimate
of the effect of the channel walls on the rotational drag by changing
the distance between the 3D printed model and the bottom of the
container. Figure~\ref{fig:walleffect} shows the relative increase in
drag when the spiral shape approaches the wall. This experiment was
performed on a \SI{5}{cm} long, \SI{5}{mm} diameter spiral at
\SI{8}{rpm}. To visualise the increase, the reciprocal of the distance
normalised to the length of the bacteria is used on the bottom horizontal
axis. The normalised length is shown on the top axes. Note that
when plotted in this way, the slope approaches unity at larger distances.

For an increase over \SI{5}{\percent}, the model has to approach the
wall at a distance smaller than $L$/3, where $L$ is the 
length of the bacteria. For very long bacteria of \SI{10}{\um}, this distance is
already reached in the middle of
the \SI{5}{\um} high channel. Since there are two channel walls on
either side at the same distance, we estimate that the additional drag
for bacteria swimming in the centre of the channel is less than
\SI{15}{\percent}. If the spiral model approaches the wall, the drag
rapidly increases. At $L/50$, the drag increases by
\SI{60}{\percent}. It is tempting to translate this effect to real
MTB. It should be noted however that the 3D printed models are rigid
and stationary, whereas the MTB are probably more flexible and
mobile. Intuitively, one might expect a lower drag.

From the bacterial dimensions, we can estimate a mean rotational drag
coefficient, $f_\text{b}$, of \SI{67(7)}{zNs}. Since the relation
between the rotational drag and the bacterial dimensions is highly
nonlinear, a Monte Carlo method was used to estimate the error and
variation of $f_\text{b}$. For these calculations, the length  of the bacteria
was assumed to be Gaussian distributed with parameters as
indicated in table~\ref{tab:MTB_torque}. The code for the Monte Carlo
calculation is available as additional material.

Due to the nonlinearity, the resulting distribution of $f_\text{b}$
is asymmetric. So rather than the standard deviation, the
\SIrange{10}{90}{\percent} cut-off values of the distribution are
given in table~\ref{tab:MTB_Calc}). Most of the MTB are estimated to
have a drag coefficient in the range of \num{30} to \SI{120}{zNs}. 

% Ik heb geen idee meer waar dit over ging, Leon
% The results suggest a biased distribution with an upper limit close to the mean of the population and a lower limit much further from the mean. Whether these are outliers or whether the population is Gaussian distributed has not been looked into, though the order of magnitude of the values obtained agree with those found in literature \cmtrmp{ref.}

\begin{table}
  \caption{Characteristics of 
    magnetotactic bacteria. Length $L$ and width
    $W$ and  amount $n$, diameter $d$ and spacing
    $a$ of the crystals in the magnetosomes. The error indicated on
    the means is the standard error (standard deviation/square root
    of the total number of samples).}
% Table is now in agreement with sagemath worksheet. If you change
% values, update the worksheet!!!!!
  \label{tab:MTB_torque}
  \begin{ruledtabular}
     \begin{tabular}{lcccccc}
      &  $L$  & $W$ & $n$ & $d$ & $a$ \\
      & [\si{\micro m}]&  [nm] &  &  [nm] & [nm]\\
      %         L       W      n       2r      a       
      mean & \num{5.0(2)} & \num{240(6)} & \num{16(2)} & \num{40(2)} & \num{56(1)}\\
      stddev  & 1 & 28 & 6 & 9 & 8
    \end{tabular}
  \end{ruledtabular}
\end{table}

% Table is now in agreement with sagemath worksheet,  update if values
% in table I change!}

\begin{table*}
  \caption{ From the values of table~\ref{tab:MTB_torque}, the  drag coefficient
    $f_\text{r}$, demagnetisation factors $\Delta N$, magnetic moment $m$, maximum magnetic torque  $\Gamma_\text{max}$, and proportionality factor
    $\gamma$ are estimated ($v/D=\gamma B$). The input parameters are
    assumed to obey a Gaussian distribution with standard deviations as in
    table~\ref{tab:MTB_torque}. Using a Monte Carlo method, the standard error
    of the calculated parameters, and the \SI{10}{\percent}--\SI{90}{\percent} cut-offs in the  distribution, are calculated.}
  \label{tab:MTB_Calc}
  \begin{ruledtabular}
    \begin{tabular}{lcccccc}
      &  $f_\text{b}$ & $\Delta N$ & $m$ & $\Gamma_\text{max}$ & $\gamma_\text{theory}$ & $\gamma_\text{exp}$\\
      &  [zNms] &  & [fAm$^2$] & [aNm]  & [rad/mTs] & [rad/mTs] \\
      %   fr      DeltaN  	 m  	Tmax     g
      mean & \num{67(7)} & \num{0.10(2)} & \num{0.25(05)} & \num{7(3)} & \num{1.2(3)} & \num{0.74(3)}\\
      10\%   & 31 & 0.03 & 0.07 & 0.7 & 0.3 & \\
      90\%   & 124 &  0.27 & 0.57 & 41 & 3.6 & \\
    \end{tabular}
  \end{ruledtabular}
\end{table*}

\subsubsection{Estimate of magnetic torque}
\label{sec:estim-magn-torq}

\begin{figure}
  \begin{centering}
    \includegraphics[width=\widefigurewidth]{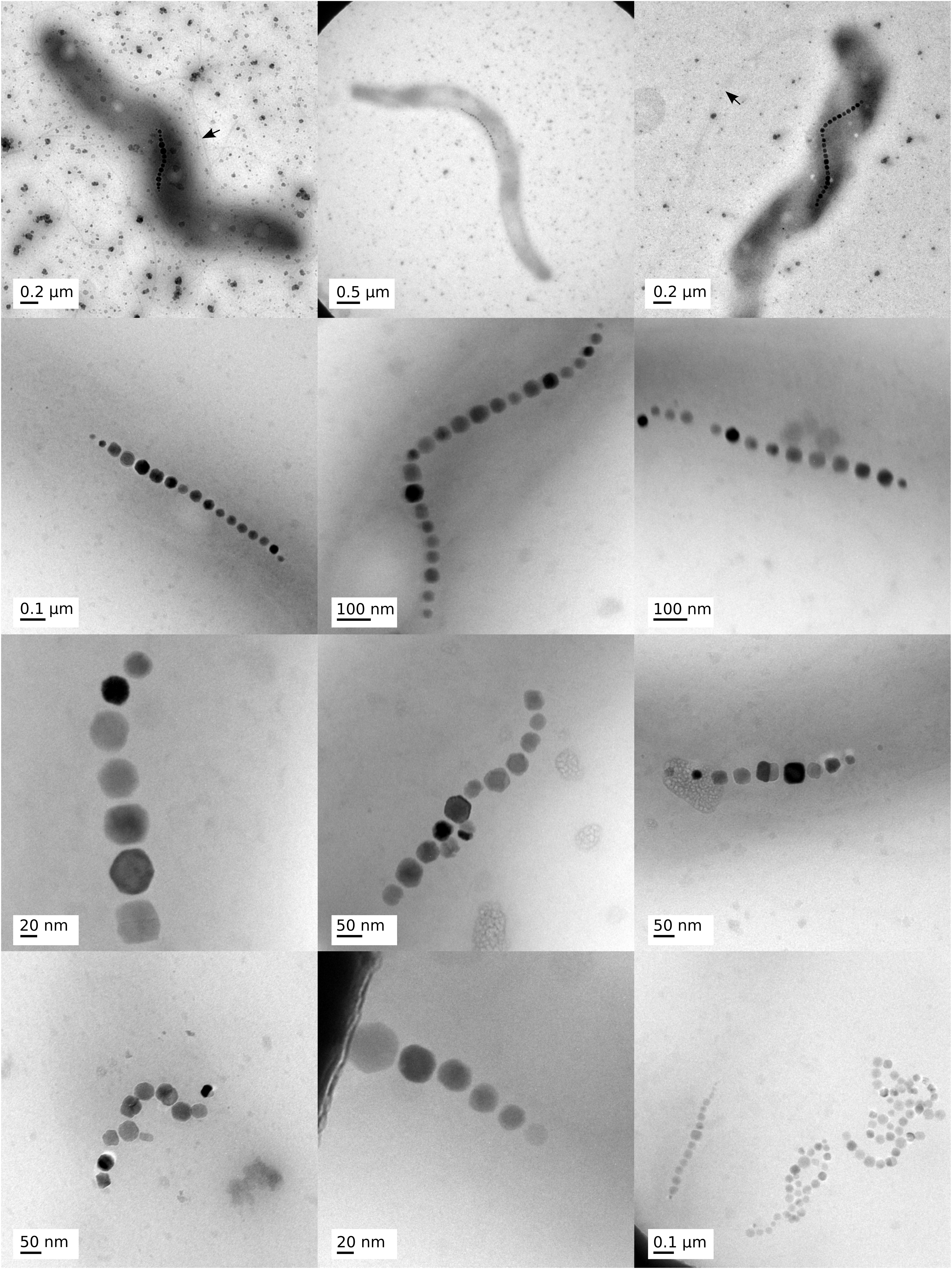}
  \end{centering}
  \caption{Transmission electron micrographs of MSR-1, magnetosomes and chains. The top row shows typical full scale bacteria, where black arrows indicate the flagella. Compared to the second row, the third row shows shorter chains with a higher variety in size distribution of magnetic nanoparticles due to an immaturity of the chain~\cite{Uebe2016}. The bottom row shows irregular chains and overlapping groups of expelled chains due to the formation of aggregates, making it hard or impossible to distinguish individual chains.}
\label{fig:MTBTEM}
\end{figure}

\begin{figure}
  \begin{centering}
    \includegraphics[width=\widefigurewidth]{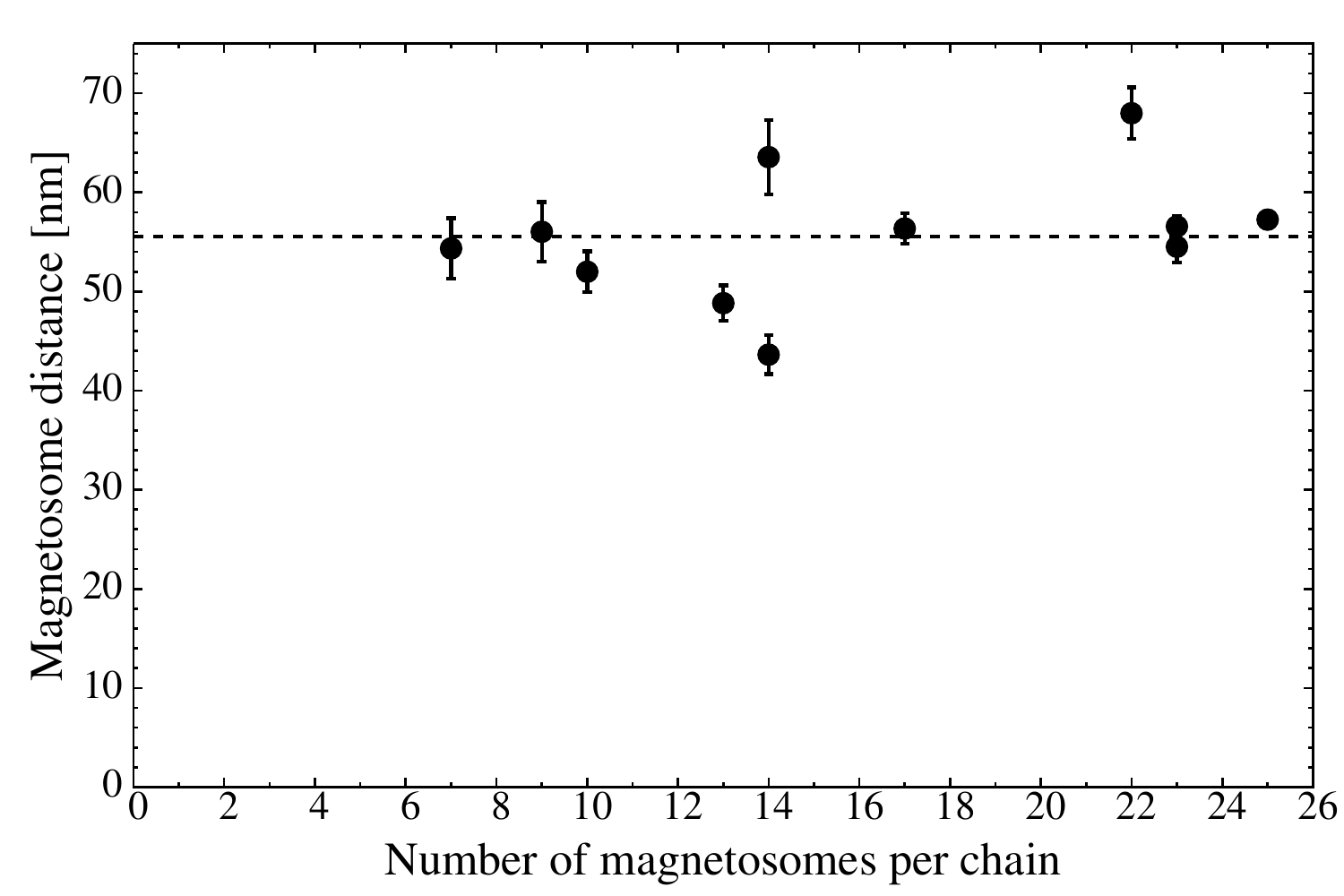}
  \end{centering}
  \caption{Distance between magnetite particles as a function of the number of particles in the chain. The mean of the entire sample group is indicated with a dashed line at \SI{56(1)}{nm}. Vertical error bars represent the standard error of each individual chain.}
\label{fig:c2clength}
\end{figure}

Figure~\ref{fig:MTBTEM} shows typical transmission electron microscopy
images (TEM) of magnetosome chains.  From these images, we obtain the
magnetosome count $n$, radius $r$, and distance $d$, which are listed
as well in table~\ref{tab:MTB_torque}. These values agree with those
reported in the literature~\cite{Posfai2007, Faivre2010} and lie within the range of single-domain magnets~\cite{Faivre2015}. We have found no
significant relation between the inter-magnetosome distance and the chain length, see
figure~\ref{fig:c2clength}.

% Wanneer klaar, discussie over verhoudingen, verschillen en overeenkomsten tussen parameters. Update nog over discrepantie tussen waarde simulaties en particles op basis van TEM (theory vs experimental)
% Since the first publication on magnetotactic bacteria (MTBs)~\cite{Blakemore1979}, a plethora of MTBs and biogenic magnetic nanoparticles (MNPs)~\cite{Devouard1998, Pankhurst2003, Majetich2003,  Schuler2008, Abhilash2010, Faivre2010} have been studied in depth for both in vivo and in vitro applications. \cmtrmp{iets hieruit halen ter vergelijking met de waarden uit onze tabellen?}

From these values the demagnetisation factor $\Delta N$, the magnetic
moment $m$, and the maximum torque $\Gamma_\text{max}$ are calculated using
the model from section~\ref{sec:angularvel}, and tabulated in
table~\ref{tab:MTB_Calc}. Again, the standard deviations of the values
and the 10\%- and \SI{90}{\percent} cut-off values are determined from
Monte Carlo simulations.

\subsubsection{Average rate of rotation}
From the drag coefficient $f_\text{r}$ and the maximum torque
$\Gamma_\text{max}$, the ratio $\gamma$ between the average rate of
rotation and the magnetic field strength can be obtained
using equation~\ref{eq:gammadefinition}. This value is listed as
$\gamma_\text{theory}$ in table~\ref{tab:MTB_Calc}, and has a
convenient value of approximately \SI{1}{rad/mTs}.  So in the earth's
magnetic field (\SI{0.04}{mT}), the rate of rotation of an MSR-1 is
approxmately \SI{0.04}{rad/s}. A U-turn will take
at least \SI{78}{s}.

\subsubsection{Average Velocity}
\begin{figure}
  \begin{centering}
    \includegraphics[width=\widefigurewidth]
    {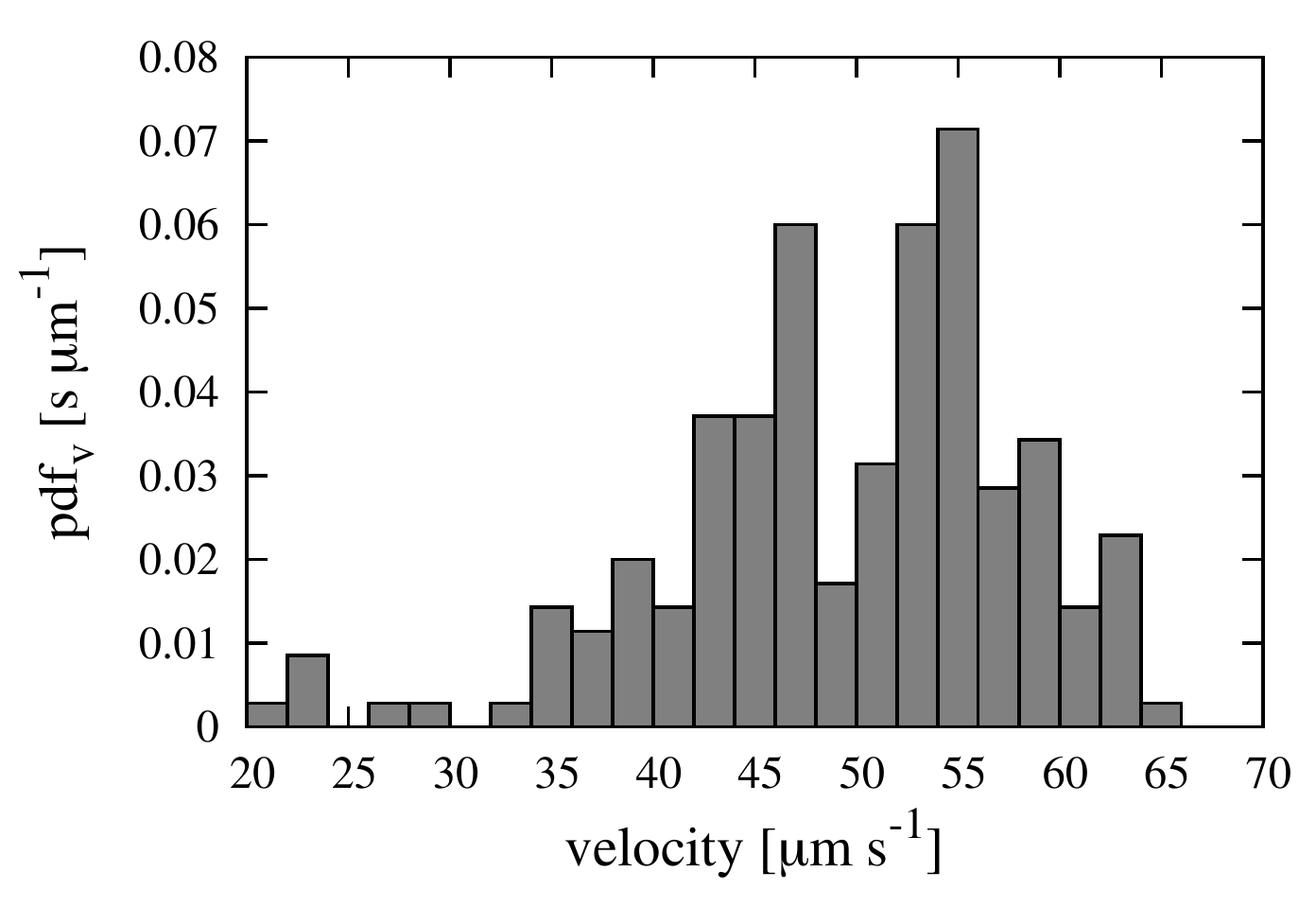}
  \end{centering}
  \caption{Probability density function for the MTB velocity
    distribution for \num{174} observed MTB.}
  \label{fig:distributionv}
\end{figure}

The MTBs' velocity was determined from the full set of 174 analyzed bacteria
trajectories. This set has a mean velocity of \SI{49.5(7)}{\um/s} with
a standard deviation $\sigma$ of \SI{8.6}{\um/s}
(figure~\ref{fig:distributionv}).
Using the value for the average rate of rotation $\gamma$ of approximately \SI{1}{rad/mTs}, this speed leads to a U-turn in the earth's magnetic
field of about \SI{1}{mm} (equation~\ref{eq:gammadefinition}).

Comparing the velocity of \emph{Magnetospirillum Gryphiswaldensen} in
the vicinity of an oxic-anoxic zone (OAZ), \SI{13(1)}{\um/s} to
\SI{23(3)}{\um/s}~\cite{Lefevre2014} (orientation towards
OAZ-dependent), or without, \SI{42(4)}{\um/s}~\cite{Popp2014},
suggests our value of the velocity of the MTB is not restricted by an oxygen gradient.
Depending on the choice of binning, one might recognise a dip in the
velocity distribution. Similar dips have been found in previous
research, which were attributed to different
swimming modes~\cite{Reufer2014}. There might as well be possible wall-effects on
bacteria caused by the restricted space in the microfluidic
chip~\cite{Magariyama2005}.

%Whether this can be attributed to a difference in velocity between MTBs performing a U-turn or simply moving in a straight line, remains unclear from our data. Therefore we assume that the velocity before, during and after U-turn reversal remains the same. The consistency in velocity will require further analysis in future research, as it will directly affect the results from the method used for calculating the average rate of rotational. \cmtrth{I would not include the previous paragraph...}
% oude velocity waardes: 50 +- 0.6 met sd van 8.2

\begin{figure}
  \begin{centering}
    \includegraphics[width=\widefigurewidth]{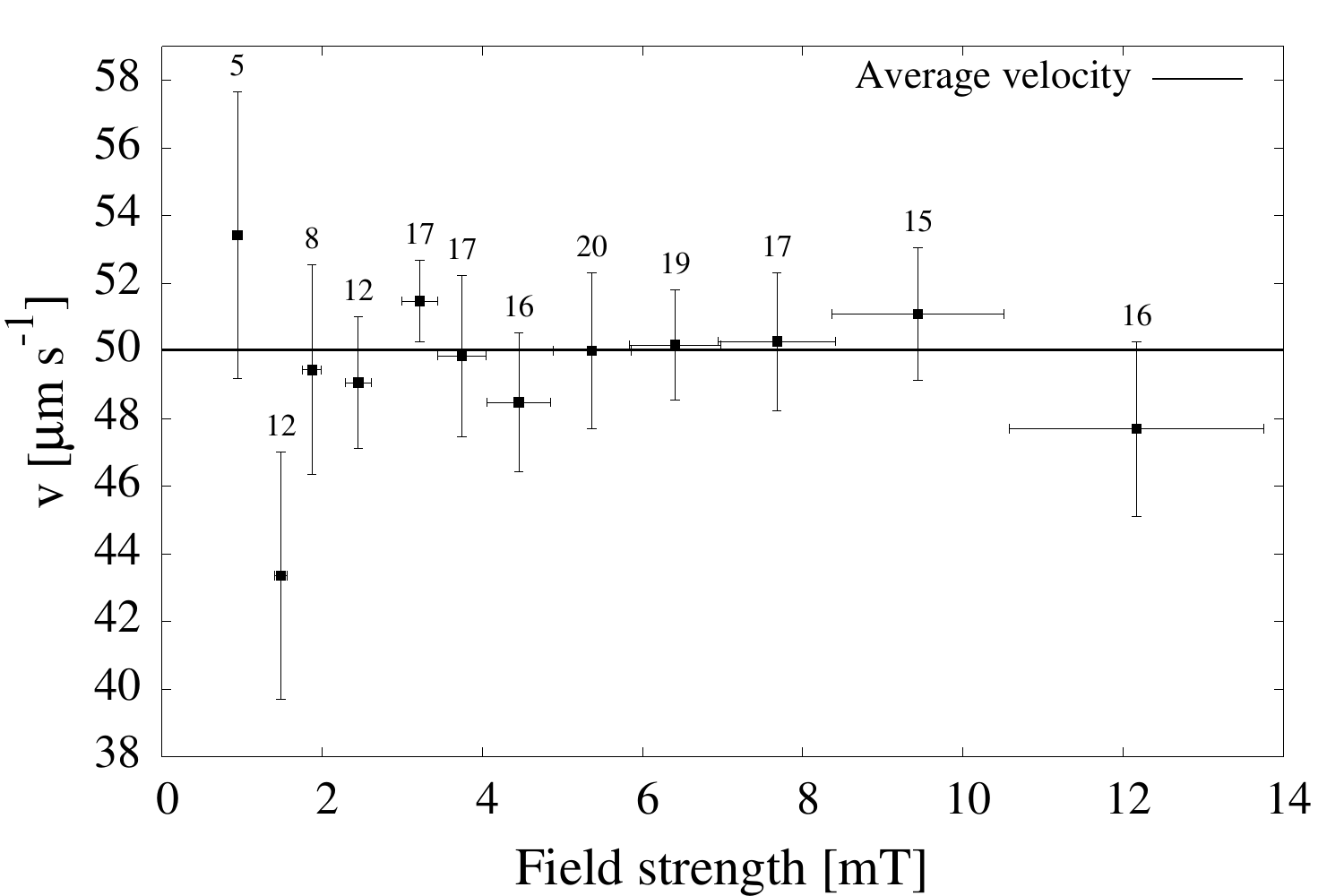}
  \end{centering}
  \caption{Average MTB velocity as function of the applied magnetic
    field. The vertical error bars indicate the standard error calculated
    from the number of bacteria indicated above the error bar.}
\label{fig:resultsv}
\end{figure}

The measured velocity during U-turns as a function of the magnetic field strength is shown in figure~\ref{fig:resultsv}. The vertical error bars display the standard error of the velocity within the group. The size of the sample group is depicted above the vertical error bars. For every sample group containing less than ten bacteria, the standard deviation of the entire population was used instead. The error in the magnetic field is due to positioning error, as described in section \ref{sec:setup}.

% The average velocity of \SI{49.5(7)}{\um/s} with a \sigma of \SI{8.6}{\um/s} uit vorige paragraaf? 
On the scale of the graph, the deviation from the mean
velocity is seemingly large, especially below \SI{2}{mT}. This deviation is
however not statistically significant. The reduced $\chi^{2}$ of the
fit to the field-independent model is very close to unity
(\num{0.67}), with a high $Q$-value of \num{0.77} (the probability
that $\chi^2$ would even exceed that value by chance, see Press~\emph{et al.}, chapter 15~\cite{Press1992}).  
%The reduced $\chi^{2}$ is the value of $\chi^{2}$ divided by the number of degrees of freedom, being in our case the number of bacteria observed (\num{174}) minus one.) \cmtr{waarom is dit gecomment?}
Within the standard errors obtained in this measurement, and for the range of field values applied, we can conclude that the velocity of the MTB is independent of the applied magnetic field, as expected.

%%% Local Variables:
%%% mode: latex
%%% TeX-master: "UTurn"
%%% End:
 
% Trajectory measurement, and comparison with model:
\subsection{Trajectories}
\label{sec:trajectories-1}
\begin{figure}
  \begin{centering}
    \includegraphics[width=\widefigurewidth]{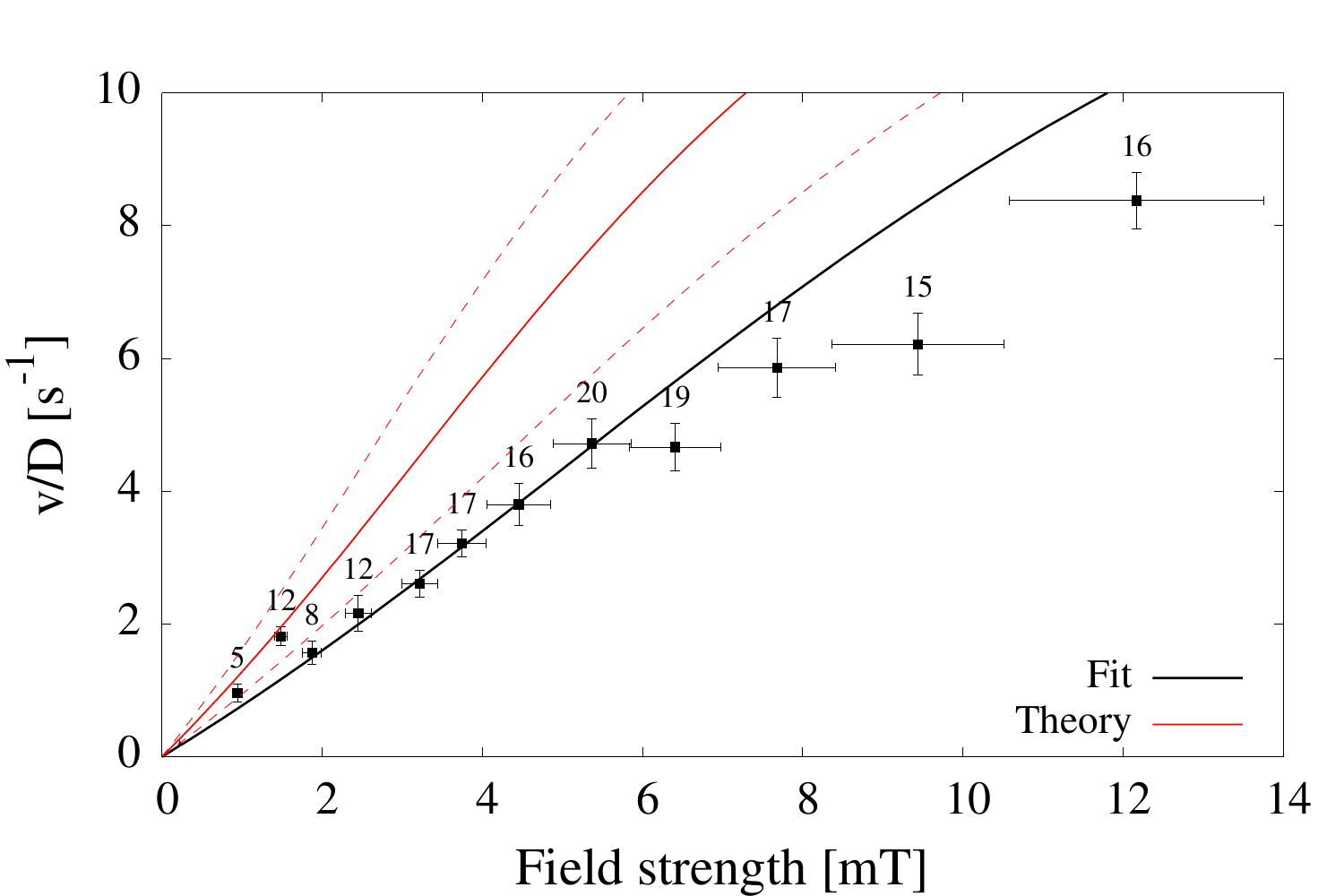}
  \end{centering}
  \caption{The average rate of rotation, $v/D$, as a function of the
    applied magnetic field, $B$. Vertical error bars display the
    standard error calculated for the number of MTB denoted above the error
    bars. For remaining sample groups, containing less than 10
    bacteria samples, the standard deviation of the entire population
    is used instead. The black solid line is the fit of the model to
    the measured data, resulting in
    $\gamma_\text{exp}=$\SI{0.74(3)}{rad/mTs}. The
    solid red line is the model prediction, using the
    $\gamma_\text{theory}$ derived from the bacteria and magnetosome
    dimensions, with the dotted red lines indicating the error on the
    estimate (\SI{1.2(3)}{rad/mTs}).}
\label{fig:resultsvr}
\end{figure}

The diameter $D$ of the U-turn was measured from the trajectories as
in figure~\ref{fig:trajectoryselect}. From these values and the measured
velocity $v$ for each individual bacterium, the average rate of rotation
$v/D$ can be calculated.  Figure~\ref{fig:resultsvr} shows this average rate of
rotation as a function of the applied magnetic field, $B$. The
error bars are defined as in figure~\ref{fig:resultsv}. 

The data points are fitted to the U-turn trajectory model simulations
of section~\ref{sec:trajectories}. The fit is shown as
a solid black line, with the proportionality factor $\gamma_\text{exp}$
equal to \SI{0.74(03)}{rad/mTs}. 
%OLD VALUE: \SI{0.84(04)}{rad/mTs}.
The reduced $\chi^{2}$ of the fit is
(\num{2.88}), and the $Q$-value (\num{0.00086})
%OLD VALUE: (\num{2.0}), and the $Q$-value (\num{0.028})

Figure~\ref{fig:resultsvr} shows that the observed average rate
of rotation in low fields is higher than the model fit in comparison with the
measurement error. We neglected the effect of the (earth's) magnetic
background field. As discussed before, at this field strength,
however, the average rate of rotation is on the order of
\SI{40}{mrad/s} and the corresponding diameter of a U-turn is on the order of
\SI{1}{mm}. The background field can therefore not be the cause of any
deviation at low field strengths. Tracking during the pre-processing step under low
fields leads to an overlap between the trajectories, which affect the
post-processing step. Due to the manual selection in the post-processing,
illustrated in figure~\ref{fig:trajectoryselect}, the preference for
uninterrupted and often shorter trajectories may have led (for lower fields) to a selection
bias to smaller curvatures. The deviation from the
linear fit below \SI{2}{mT} could therefore be attributed to human
bias (``cherry picking'').
 
If we neglect trajectories below \SI{2}{mT} for this reason, the fits
improve (drastically) for both the velocity and average rate of
rotation. Fitting datapoints over the range of \SIrange{2}{12} mT (eight
degrees of freedom) decreases the reduced $\chi^{2}$ of the velocity
from \num{0.67} to \num{0.42}. Furthermore, the $Q$-value of \num{0.77}
is increased to \num{0.91}, a slight increase in likelihood that our
datapoints fall within the limits of the model.

%OLD VALUES: Similarly, the reduced $\chi^{2}$ of the average rate of rotation is lowered from \num{2.0} to \num{0.69} and the %$Q$-value from \num{0.03} to \num{0.70}, a drastic change in likelihood of the fit. We therefore assume that these %results validate the model with the exclusion of outliers below \SI{2}{mT}.
Similarly, the reduced $\chi^{2}$ of the average rate of rotation is lowered from \num{2.88} to \num{1.03} and the $Q$-value from \num{0.00086} to \num{0.41}, a drastic change in likelihood of the fit. We therefore assume that these results validate the model with the exclusion of outliers below \SI{2}{mT}.

At high fields, the observed average rate of rotation seems to be on
the low side, although within the error bounds. For the high field range,
the diameter of a U-turn is on the order of \SI{5}{\um} and reversal times
are on the order of \SI{100}{ms}. The resolving power of our setup of
\SI{180}{\nm/pixel} and time resolution of \SI{100}{frames/s} are
sufficient to capture these events, so cannot explain the apparent
deviation. It is possible that the weakest bacteria reach the
saturation torque (figure~\ref{fig:MTBTorqueVersusField}), although
the effect is not expected to be very significant.

%%% Local Variables:
%%% mode: latex
%%% TeX-master: "UTurn"
%%% End:

%%% Local Variables:
%%% mode: latex
%%% TeX-master: "UTurn"
%%% End:

\section{Discussion}
\label{sec:discussion}

Figure~\ref{fig:resultsvr} shows in red the prediction of the model using the proportionality factor determined from observations of the MTB
(the outer dimensions by optical microsopy and SEM,
the magnetosome by TEM), $\gamma_\text{theory}$=\SI{1.2(3)}{rad/mTs}. The
predicted proportionaliy factor is clearly higher than measured. This is
either because we overestimated the magnetic moment or underestimated
the rotational drag coeffient. The latter seems more likely. In the
the first place, we neglected the influence of the flagella. A coarse
estimate using a rigid cylinder model for the flagellum shows that a
flagellum could indeed cause this type of increase in drag. Since we
lack information on the flexibility of the flagellum, we cannot
quantify the additional drag. Secondly, we ignored the finite height
of the microfluidic channel. As was shown by the macroscale
experiments, the additional drag increases rapidly if a bacterium
approaches within a few hundred nanometers of the wall. Since we do not have
information about the distance, again quantification is difficult.

Given the above considerations, we are confident that over
the observed field range, the MTB trajectories are in fair agreement
with our model.

\section{Conclusion}
\label{sec:conclusion}
% conclusies TEM en Trajectories

% Onderzoeksvraag
We studied the response of the magnetotactic bacteria
\emph{Magnetospirillum Gryphiswaldense} to rotation of an external
magnetic field $B$, ranging in amplitude from \SI{1}{mT} up to
\SI{12}{mT}.

% Conclusies, theorie en experimental
Our magnetic model shows that the torque on the MTB is linear in the
applied field up to \SI{10}{mT}, after which the torque starts to
saturate for an increasing part of the population.

Our theoretical analysis of bacterial trajectories shows that the
bacteria perform a U-turn under \SI{180}{\degree} rotation of the
external field, but not at a constant angular velocity. The 
diameter, $D$, of the U-turn increases with an increase in the velocity $v$ of the bacteria. The average rate
of rotation, $v/D$, for an instantaneously reversing field is linear
within \SI{2}{\%} in the applied field up to \SI{12}{mT}.

If the applied field is rotated over \SI{180}{\degree} in a finite
time, the average rate of rotation is higher at low field values
than it was for an instantaneous reversal. Given a
field rotation time, an optimum field value exist at which the rate of
rotation is approximately \SI{18}{\percent} higher than for
instantaneous reversal. This optimum field value is inversely
proportional to the field rotation time.

% Macroscopic experiments
The rotational drag coefficient for an MTB was estimated from drag
rotation experiments in a highly viscous fluid, using a macroscale 3D
printed MTB model. The spiral shape of the body of an MTB has a
\SI{64(5)}{\percent} higher drag than a spheroid with equal length
and diameter, which has been the default model in the literature up to
now. Furthermore, the added drag from the channel wall was found to be
negligible for an MTB in the center between the walls (less than
\SI{10}{\percent}), but to increase rapidly when the MTB approaches to within a few hundred nanometers of one
of the walls. 

% Microscopic
From microscope observations, we conclude that the MTB velocity during
a U-turn is independent of the applied field. The population of MTB
has a non-Gaussian distributed velocity, with an average of
\SI{49.5(7)}{\um/s} and a standard deviation of \SI{8.6}{\um/s}. As
predicted by our model, the average rate of rotation is linear in
the external magnetic field within the measured range of
\SIrange{1}{12}{mT}. The proportionality factor $\gamma=v/DB$ equals
\SI{0.74(3)}{rad/mTs}. The predicted theoretical value
is \SI{1.2(3)}{rad/mTs}, which is based on measurements of the
parameters needed for the model, such as the size of the bacteria and
their magnetosomes from optical microscopy, SEM, and TEM images. The
number of parameters and their nonlinear relation with the
proportionality factor causes the relatively large error in the
estimate.

These findings finally prove that the generally accepted linear model
for the response of MTB to external magnetic fields is correct within the
errors caused by the estimation of the model parameters if the field values are
below \SI{12}{mT}. At higher values, torque saturation will occur. 

This result is of importance to the control engineering community. The
knowledge of the relation between the angular velocity and the field
strength ($\gamma$) can be used to design energy efficient control
algorithms that prevent the use of excessive field
strengths. Furthermore, a better understanding of the magnetic
behavior will lead to more accurate predictions of the dynamic
response of MTB for potential applications in micro-surgery, as drug
carriers, or for drug delivery.

%%% Local Variables:
%%% mode: latex
%%% TeX-master: "UTurn"
%%% End:

\begin{acknowledgments}
  The authors wish to thank Matthias Altmeyer of KIST Europe for
  assistance in cultivating the bacteria, Lars Zondervan of the
  University of Twente for calculations of demagnetization factors,
  Jorg Schmauch of the Institute for New Materials in Saarbr\"ucken, Germany,
  for TEM imaging, and Carsten Brill of KIST Europe for SEM
  imaging.
\end{acknowledgments}

% Make symbolic links to the PaperBase directory. 
% Make sure you checkout out the svn PaperBase directory
% Open terminal
% cd <your svn path>/Marc/Uturn
% ln -s ../../PaperBase/paperbase.bib .
% ln -s ../../PaperBase/bst .
\bibliographystyle{bst/apsrev_modified_doi}
\bibliography{paperbase}
\end{document}